\documentclass[apj,numberedappendix]{emulateapj}
\usepackage{apjfonts}

\usepackage{lscape}
\usepackage{epsf}

\defcitealias{fall01}{FZ01}
\defcitealias{baumgardt03}{BM03}

\slugcomment{The Astrophysical Journal, 679:1272--1287, 2008 June 1}

\shorttitle{SHAPING THE GCMF BY EVAPORATION}
\shortauthors{McLAUGHLIN \& FALL}

\begin{document}

\title{Shaping the Globular Cluster Mass Function by Stellar-Dynamical
       Evaporation}

\author{Dean E. McLaughlin\altaffilmark{1,2} and
        S. Michael Fall\altaffilmark{3,4}}

\altaffiltext{1}{Dept.~of Physics and Astronomy, University of Leicester,
       University Road, Leicester, UK LE1 7RH}
\altaffiltext{2}{Permanent address: Astrophysics Group,
       Lennard-Jones Laboratories, Keele University, Keele,
       Staffordshire, UK ST5 5BG; {\tt dem@astro.keele.ac.uk}}
\altaffiltext{3}{Institute for Advanced Study, Einstein Drive,
       Princeton, NJ 08450}
\altaffiltext{4}{Permanent address: Space Telescope Science Institute,
       3700 San Martin Drive, Baltimore, MD 21218;
       {\tt fall@stsci.edu}} 

\begin{abstract}

We show that the globular cluster mass function (GCMF) in the Milky Way
depends on cluster half-mass density, $\rho_h$, in the sense that the
turnover mass $M_{\rm TO}$ increases with $\rho_h$ while the width of
the GCMF decreases. We argue that this is the expected signature of
the slow erosion of a mass function that initially rose towards low
masses, predominantly through cluster evaporation driven by internal
two-body relaxation. We find excellent agreement between the observed
GCMF---including its dependence on internal density $rho_h$, central 
concentration $c$, and Galactocentric distance $r_{\rm gc}$---and a
simple model in which the relaxation-driven mass-loss rates of clusters
are approximated by $-dM/dt = \mu_{\rm ev} \propto \rho_h^{1/2}$. In
particular, we recover the well-known insensitivity of $M_{\rm TO}$ to
$r_{\rm gc}$. This feature does not derive from a literal
``universality'' of the GCMF turnover mass, but rather from a
significant variation of $M_{\rm TO}$ with $\rho_h$---the expected
outcome of relaxation-driven cluster disruption---plus significant
scatter in $\rho_h$ as a function of $r_{\rm gc}$. Our conclusions are
the same if the evaporation rates are assumed to depend instead on the
mean volume or surface densities of clusters inside their tidal radii,
as $\mu_{\rm ev} \propto \rho_t^{1/2}$ or
$\mu_{\rm ev} \propto \Sigma_t^{3/4}$---alternative prescriptions that
are physically motivated but involve cluster properties ($\rho_t$ and
$\Sigma_t$) that are not as well defined or as readily observable as
$\rho_h$. In all cases, the normalization of  $\mu_{\rm ev}$ required
to fit the GCMF implies cluster lifetimes that are within the range of
standard values (although falling towards the low end of this
range). Our analysis does not depend on any assumptions or information
about velocity anisotropy in the globular cluster system.

\end{abstract}


\section{Introduction}
\label{sec:intro}

The mass functions of star cluster systems provide an important point
of reference for 
attempts to understand the connection between old globular clusters
(GCs) and the young massive clusters that form in local starbursts and
galaxy mergers. When expressed as the number per unit logarithmic mass,
$dN/d\,\log\,M$, the GC mass function (GCMF) is characterized by a
peak, or turnover, at a mass
$M_{\rm TO} \approx 1$--$2\times 10^5\,M_\odot$ that is empirically
very similar in most galaxies. By contrast, the mass functions of
young clusters show no such feature but instead rise
monotonically towards low masses over the full observed range
($10^6\,M_\odot \ga M \ga 10^4\,M_\odot$ in the best-studied cases),
in a way that is well described by a power law,
$dN/d\,\log\,M \propto M^{1-\beta}$ with 
$\beta \simeq 2$ \citep[e.g.,][]{zhang99}.

At the same time, for high
$M > M_{\rm TO}$, old GCMFs closely resemble the mass
functions of young clusters, and of molecular clouds in the
Milky Way and other galaxies \citep{harris94,elmegreen97}; and
it is well known that a number of dynamical processes
cause star clusters to lose mass and can lead to their complete
destruction as they orbit for a Hubble time in the potential wells of
their parent galaxies
\citep[e.g.,][]{fall77,caputo84,aguilar88,chernoff90,gnedost97,murali97}.
It is therefore natural to ask whether the peaks in GCMFs can
be explained by the depletion over many Gyr of globulars from initial
mass distributions that were similar to those of young clusters below
$M_{\rm TO}$ as well as above.

Our chief purpose in this paper is to establish and interpret an
aspect of the Galactic GCMF that appears fundamental but has gone
largely unnoticed to date: $dN/d\,\log\,M$ has a strong and 
systematic dependence on GC half-mass density,
$\rho_h \equiv 3M/8\pi r_h^3$ ($r_h$ being the cluster half-mass
radius), in the sense that the turnover mass $M_{\rm TO}$ increases
and the width of the distribution decreases with increasing
$\rho_h$. As observed facts, these must be explained by any theory  
of the GCMF.
We argue here that they are an expected signature of slow dynamical
evolution from a mass function that initially increased
towards $M < M_{\rm TO}$, if the long-term mass loss from surviving GCs
has been dominated by stellar escape due to internal, two-body
relaxation (which we refer to from now on as either relaxation-driven 
evaporation or simply evaporation). 

\citeauthor{fall01} (\citeyear{fall01}; hereafter \citetalias{fall01})
explain in detail why cluster evaporation
dominates the long-term evolution of the low-mass shape of
observable GCMFs. Briefly,
stellar evolution removes (through supernovae and winds) the
same {\it fraction} of mass from all clusters of a given age, and so
cannot change the shape of $dN/d\,\log\,M$ (unless special initial
conditions are invoked; cf.~\citealt{vesperini03a}).
Meanwhile, for GCs like those that have survived for a Hubble
time in the Milky Way, the mass loss from gravitational shocks
during disk crossings and bulge passages is generally less than
that due to evaporation for $M<M_{\rm TO}$ (\citetalias{fall01};
\citealt{prieto06}).\footnotemark
\footnotetext{It is possible that there existed a past population of
  GCs with low densities or concentrations, or perhaps on extreme
  orbits, that were destroyed in less than a Hubble time by shocks
  or stellar evolution. Our discussion does not cover such clusters.}

As we discuss further in \S\ref{sec:results} below,
the evaporation of tidally limited clusters proceeds at a rate,
$\mu_{\rm ev} \equiv -dM/dt$, that is approximately constant in time
and primarily determined by cluster density.
\citetalias{fall01} show that a constant mass-loss rate
leads to a power-law scaling $dN/d\,\log\,M \propto
M^{1-\beta}$ with $\beta \rightarrow 0$ (corresponding to a flat
distribution of clusters per unit {\it linear} mass) at sufficiently 
low $M < \mu_{\rm ev} t$ in the evolved mass function of
coeval GCs that began with {\it any} nontrivial initial
$dN/d\,\log\,M_0$.\footnotemark 
\footnotetext{Throughout this paper, we use ``initial'' to mean at a
  relatively early time in the development of long-lived clusters,
  after they have dispersed any remnants of their natal gas clouds,
  survived the bulk of stellar-evolution mass loss, and come into
  virial equilibrium in the tidal field of a galaxy.} 
To accommodate this when $dN/d\,\log\,M_0$ originally increased
towards low masses as a power law, a time-dependent peak must develop
in the GCMF at a mass of order $M_{\rm TO}(t) \sim \mu_{\rm ev} t$
(\citetalias{fall01}). But then, since $\mu_{\rm ev}$ depends
fundamentally on cluster density, so too must $M_{\rm TO}$.

A $\beta \simeq 0$ power-law scaling below the turnover mass has
been confirmed directly in the GCMFs of the Milky Way
(\citetalias{fall01}) and the giant elliptical M87 \citep{waters06},
while \citet{jordan07} show it to be consistent with
$dN/d\,\log\,M$ data for 89 Virgo Cluster galaxies, and it is
apparent in deep observations of some other GCMFs
(e.g., in the Sombrero galaxy, M104; \citealt{spitler06}).
As regards the peak itself, old GCs are observed
\citep[e.g.,][]{jordan05} to have rather similar densities on
average---and, therefore, similar typical $\mu_{\rm ev}$---in galaxies
with widely different total luminosities and Hubble types.
(Inasmuch as cluster densities are set by tides, this is probably
related to the mild variation of mean {\it galaxy} density with total
luminosity; see \citetalias{fall01}, and also
\citealt{jordan07}.) Thus, an evaporation-dominated evolutionary
origin for a turnover in the GCMF appears to be consistent with the
well-known fact that the mass scale $M_{\rm TO}$ generally differs
very little among galaxies \citep[e.g.,][]{harris01,jordan06}.

If this picture is basically correct, it implies that,
even though $M_{\rm TO}$ may appear nearly universal when considering
the global mass functions of entire GC systems, in fact the GCMFs of
subsamples of clusters with similar ages but different densities
should have different turnovers. In \S\ref{sec:results}, we
show---working for definiteness and relatively easy observability with
the half-mass density, $\rho_h$---that this is the case for globulars
in the Milky Way. We fit the observed $dN/d\,\log\,M$ for GCs in bins of
different $\rho_h$ with models assuming that
(1) the initial distribution increased as a $\beta=2$ power law at low
masses and 
(2) the mass-loss rates of individual clusters can be estimated
from their half-mass densities by the rule
$\mu_{\rm ev} \propto \rho_h^{1/2}$.
In \S\ref{sec:discussion} we discuss the validity of this prescription
for $\mu_{\rm ev}$, which is certainly approximate but captures the
main physical dependence of relaxation-driven mass loss.
In particular, we show that the alternative mass-loss laws
$\mu_{\rm ev} \propto \rho_t^{1/2}$ and
$\mu_{\rm ev} \propto \Sigma_t^{3/4}$---where $\rho_t$ and $\Sigma_t$
are the mean volume and surface densities inside cluster tidal
radii---lead to models for the GCMF that are
essentially indistinguishable from those based on
$\mu_{\rm ev} \propto \rho_h^{1/2}$.
The normalization of $\mu_{\rm ev}$ required to fit the observed GCMF
implies cluster lifetimes that are within a 
factor of $\approx\!2$ (perhaps slightly on the low side, if the
initial power-law exponent at low masses was $\beta=2$) of typical
values in theories and simulations of two-body relaxation in tidally
limited GCs.

We also show in \S\ref{sec:results} that when the observed densities
of individual clusters are used in our models to predict GCMFs in
different bins of Galactocentric radius ($r_{\rm gc}$), they fit the
much weaker variation of $dN/d\,\log\,M$ and $M_{\rm TO}$ as functions
of $r_{\rm gc}$, which is well-known in the Milky Way
and other large galaxies
\citep[see][]{harris01,harris98,barmby01,vesperini03b,jordan07}. 
Similarly, applying our models to the GCs in two bins of central
concentration, with only the measured $\rho_h$ of the clusters in each
subsample as input, suffices to account for previously noted
differences between the mass functions of low- and high-concentration
Galactic globulars \citep{smith02}. The most
fundamental feature of the GCMF therefore appears to be its dependence
on cluster density, which can be understood at least qualitatively
(and even quantitatively, to within a factor of 2)
in terms of evaporation-dominated cluster disruption.

There is a widespread perception that if the GCMF evolved slowly
from a rising power law at low masses, then a weak or
null variation of $M_{\rm TO}$ with $r_{\rm gc}$ can be achieved only
in GC systems with strongly radially anisotropic velocity
distributions, which are not observed (see especially
\citealt{vesperini03b}).
This apparent inconsistency has been cited to bolster some recent
attempts to identify a mechanism by which a ``universal'' peak at
$M_{\rm TO} \sim 10^5\,M_\odot$ might have been imprinted on the GCMF
at the time of cluster formation, or very shortly afterwards, and
little affected by the subsequent destruction of lower-mass GCs
\citep[e.g.,][]{vesperini03a,parmentier07}.
However, given the real successes of an evaporation-dominated
evolutionary scenario for the origin of $M_{\rm TO}$, as summarized
above and added to below, it would be premature to reject the
idea in favor of {\it requiring} a near-formation origin, solely on
the basis of difficulties with GC kinematics.
(And, in any event, formation-oriented models must now be reconsidered
in light of the {\it non}-universality of $M_{\rm TO}$ as a function
of cluster density.)

We are not concerned in this paper with velocity anisotropy in GC
systems, because we only predict an 
evaporation-evolved $dN/d\,\log\,M$ as a function of cluster density
(and age) and take the observed distribution of $\rho_h$ versus 
$r_{\rm gc}$ in the Milky Way as a given, to show consistency with the
observed behavior of $M_{\rm TO}$ as a function of $r_{\rm gc}$.
Most other models (\citetalias{fall01}; \citealt{vesperini03b};
and references therein) predict dynamically evolved GCMFs directly in
terms of $r_{\rm gc}$, and in doing so are forced also to derive
theoretical dependences of cluster density on $r_{\rm gc}$. It is only
at this stage that GC orbital distributions enter the problem, and
then only in conjunction with several other assumptions and
simplifications. As we discuss further in \S\ref{sec:discussion}
below, the radially biased GC velocity distributions that
appear in such models could well be consequences of one or more of
these other assumptions, rather than of the main hypothesis about
evaporation-dominated GCMF evolution.

\section{The Galactic GCMF as a Function of Cluster Density}
\label{sec:results}

In this section we define and model the dependence of the Galactic
GCMF on cluster density. First, we describe the dependence
that is expected to arise from evaporation-dominated evolution.

Two-body relaxation in a tidally limited GC leads to a roughly steady
rate of mass loss, $\mu_{\rm ev} \equiv -dM/dt \simeq {\rm constant}$
in time. Thus, the total cluster mass decreases approximately
linearly, as $M(t) \simeq M_0 - \mu_{\rm ev} t$.
This behavior is exact in some classic models of GC evolution
\citep{henon61} and is found to be a good approximation in most other
calculations (e.g., \citealt{lee87,chernoff90,vesperini97b,gnedin99,
baumgardt01,giersz01,baumgardt03,trenti07}). The result comes from a
variety of computational methods (semi-analytical, Fokker-Planck,
Monte Carlo, and $N$-body simulation) applied to clusters with
different initial conditions (densities and concentrations) on
different kinds of orbits (circular and eccentric; with and without
external gravitational shocks) and with different internal processes
and ingredients (with or without stellar mass spectra,
binaries, and central black holes). To be sure, deviations
from perfect linearity in $M(t)$ do occur, but these are generally
small---especially away from the endpoints of the 
evolution, i.e., for $0.9\ga M(t)/M_0\ga 0.1$---and neglecting them to
assume an approximately constant $dM/dt$ is entirely appropriate for our
purposes.

When gravitational shocks are subdominant to relaxation-driven
evaporation, as they generally appear to be for extant GCs, they work
to boost the mass-loss rate $\mu_{\rm ev}$ slightly without altering
the basic linearity of $M(t)$ (e.g., \citealt{vesperini97b};
\citealt{gnedin99}; see also Figure 1 of \citetalias{fall01}). 
A time-dependent mass scale $\Delta \equiv \mu_{\rm ev} t$ is then
associated naturally with any system of coeval clusters having a common
mass-loss rate: all those with initial $M_0 \le \Delta$ are
disrupted by time $t$, and replaced with the remnants of
objects that began with $M_0 > \Delta$. As we mentioned in
\S\ref{sec:intro}, if the initial GCMF increased
towards low masses as a power law, then $\Delta$ is closely related to
a peak in the evolved distribution, which eventually decreases towards
low $M<\Delta$ as $dN/d\,\log\,M \propto M^{1-\beta}$ with 
$\beta = 0$ (\citetalias{fall01}). 

In standard theory (e.g., \citealt{spitzer87}; \citealt{bt87},
Section 8.3), the lifetime of a cluster against evaporation
is a multiple of its two-body relaxation time, $t_{\rm rlx}$. For
a total mass $M$ of stars within a radius $r$, this scales to
first order (ignoring a weak mass dependence in the Coulomb
logarithm) as
$t_{\rm rlx}(r) \propto (M r^3)^{1/2} \propto M/\rho^{1/2}$, where
$\rho \propto M/r^3$. In a concentrated cluster with an internal
density gradient, $t_{\rm rlx}(r)$ of course varies throughout
the cluster, and the global relaxation timescale is an
average of the local values (see the early
discussion by \citealt{king58}). This can still be written as
$t_{\rm rlx} \propto M/\overline{\rho}^{1/2}$, with $M$ the
total cluster mass and $\overline{\rho}$ an appropriate
reference density. We then have for the instantaneous mass-loss rate,
$\mu_{\rm ev} \equiv -dM/dt \propto M/t_{\rm rlx}
                            \propto \overline{\rho}^{1/2}$.
Insofar as this is approximately constant in time, a GCMF evolving
from an initial $\beta > 1$ power law at low masses should therefore
develop a peak at a mass that depends on cluster density and age
through the parameter $\Delta \propto \overline{\rho}^{1/2} t$.

It remains to identify the best measure of $\overline{\rho}$ in this
context. A standard choice in the literature, and the one that we
eventually make to derive our main results in this paper, is the
half-mass density $\rho_h = 3M/8 \pi r_h^3$. However, in a steady
tidal field, the mean density $\rho_t$ inside the tidal radius of a
cluster is constant by definition, and thus choosing
$\overline{\rho} = \rho_t$ instead is the simplest way to ensure
that $\mu_{\rm ev} \propto \overline{\rho}^{1/2}$ and 
$\mu_{\rm ev} \simeq {\rm constant}$ in time are mutually consistent.
In fact, \citet{king66} found from direct calculations
of the escape rate at each radius within his standard (lowered
Maxwellian) models, that the coefficient in
$\mu_{\rm ev} \propto \rho_t^{1/2}$ 
is only a weak function of the internal density
structure (concentration) of the models, and thus only a
weak function of time for a cluster evolving quasistatically through a
series of such models.

The rule $\mu_{\rm ev} \propto \rho_t^{1/2}$ is routinely used to set
the GC mass-loss rates in models for the dynamical evolution of the
GCMF, although such studies normally express $\mu_{\rm ev}$
immediately in terms of orbital pericenters, $r_p$, most often by
assuming $\rho_t \propto r_p^{-2}$ as for GCs in galaxies whose total
mass distributions follow a singular isothermal sphere (e.g.,
\citealt{vesperini97a,vesperini98,vesperini00,vesperini01,vesperini03b}; 
\citealt{baumgardt98}; \citetalias{fall01}). This bypasses any explicit
examination of the GCMF as a function of cluster density, which is our
main goal in this paper. But it is done in part because tidal radii
are the most 
poorly constrained of all structural parameters for GCs in the Milky
Way (their theoretical definition is imprecise and their empirical
estimation is highly model-dependent and sensitive to low-surface
brightness data), and they are exceedingly difficult if not
impossible to measure in distant 
galaxies. We deal with this here by focusing on the GCMF as a function
of cluster density $\rho_h$ inside the less ambiguous, empirically
better determined and more robust half-mass radius, asking
how simple models with $\mu_{\rm ev} \propto \rho_h^{1/2}$ fare
against the data.

Taking $\mu_{\rm ev} \propto \rho_h^{1/2}$ in place of
$\mu_{\rm ev} \propto \rho_t^{1/2}$, which we do to construct
evaporation-evolved model GCMFs in \S\ref{subsec:models}, is most
appropriate if the ratio $\rho_t/\rho_h$ is the same for all clusters
and constant in time. This is the case in \citeauthor{henon61}'s
(\citeyear{henon61}) model of GC evolution, and in this limit
(adopted by \citetalias{fall01} in their models for
the Galactic GCMF) our analysis is rigorously justified.
However, real clusters are not homologous ($\rho_t/\rho_h$ differs
among clusters) and they do not evolve self-similarly ($\rho_h$ may
vary in time even if $\rho_t$ does not). The key assumption in our
models is that $\mu_{\rm ev}$ is approximately independent 
of time for any GC, which is well-founded in any case. By using
current $\rho_h$ values to 
estimate $\mu_{\rm ev}$, we do {\it not} suppose that the half-mass
densities are also constant, but we in effect use a single number for
all GCs to represent a range of $(\rho_t/\rho_h)^{1/2}$. Equivalently,
we ignore a dependence on cluster concentration in the normalization
of $\mu_{\rm ev} \propto \rho_h^{1/2}$. As we discuss further in
\S\ref{sec:discussion}, it is reasonable to neglect this complication
in a first approximation because $(\rho_t/\rho_h)^{1/2}$ varies
much less among Galactic globulars than $\rho_t$ and $\rho_h$ do
separately. We demonstrate this explicitly by repeating our analysis
with $\rho_h$ replaced by $\rho_t$ and recover essentially the same
results for the GCMF.

In \S\ref{sec:discussion} we also discuss some recent results, which
indicate that the timescale for relaxation-driven evaporation depends
on a slightly less-than-linear power of $t_{\rm rlx}$
(\citealt{baumgardt01,baumgardt03}). We point out that this
implies that $\mu_{\rm ev}$ may increase as a modest power
of the average {\it surface} density of a cluster as well as (or, in
an important special case, instead of) the usual volume
density. However, we show in detail that making the appropriate
changes throughout the rest of the present section to reflect this
possibility does not change any of our conclusions.

\subsection{Data}
\label{subsec:data}

\begin{figure*}
\centerline{\hfil
   \rotatebox{270}{\includegraphics[height=175mm]{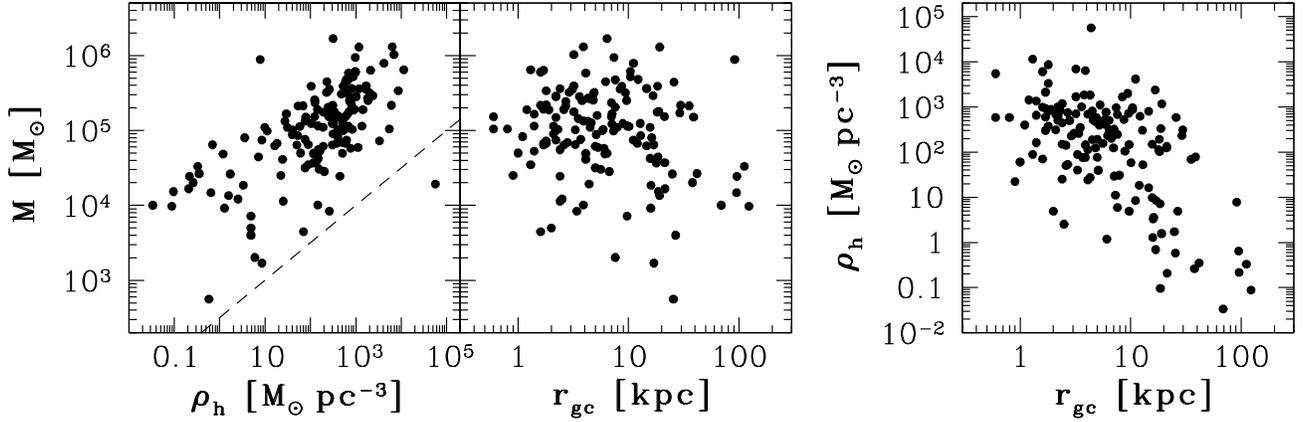}}
\hfil}
\caption{
  {\it Left:} Mass versus three-dimensional half-mass density,
  $\rho_h\equiv 3M/8\pi r_h^3$, and versus Galactocentric
  radius, $r_{\rm gc}$, for 146 Milky Way GCs in the catalogue of
  \citet{harris96}. The dashed line in the first panel is 
  $M\propto \rho_h^{1/2}$, a locus of approximately constant lifetime
  against evaporation. 
  {\it Right:} Half-mass density versus $r_{\rm gc}$ for the same clusters.
\label{fig:rhoh}}
\end{figure*}

Figure \ref{fig:rhoh} shows the distribution of mass against half-mass
density and against Galactocentric radius for 146 Milky Way GCs in
the catalogue of \citet{harris96},\footnotemark
\footnotetext{Feb.~2003 version; see
  http://physwww.mcmaster.ca/$\sim$harris/mwgc.dat \ .}
along with the distribution of $\rho_h$ versus $r_{\rm gc}$ linking
the two mass plots. The \citeauthor{harris96} catalogue actually records
the absolute $V$ magnitudes of the GCs. We obtain masses from these by
applying the population-synthesis model mass-to-light ratios
$\Upsilon_V$ computed by \citet{mcl05} for individual clusters based
on their metallicities and an assumed age of 13 Gyr. However, we 
first multiplied all of the \citeauthor{mcl05} $\Upsilon_V$ values
by a factor of 0.8 so as to obtain a median
$\widehat{\Upsilon}_V \simeq 1.5\,M_\odot\,L_\odot^{-1}$
in the end,\footnotemark
\footnotetext{Throughout this paper, we use $\widehat{x}$ to denote
  the median of any quantity $x$.}
consistent with direct dynamical estimates
(see \citealt{mcl00} and \citealt{mcl05}; also \citealt{barmby07}).

By assigning mass-to-light ratios to GCs in this
way, we allow for expected differences between clusters
with different metallicities. Our application of a corrective factor
to the population-synthesis values, $\Upsilon_V^{\rm pop}$, is
motivated empirically by the fact that their distribution among
Galactic GCs is strongly peaked around a median 
$\widehat{\Upsilon}_V^{\rm pop}\simeq 1.9\ M_\odot\,L_\odot^{-1}$,
while the observed (dynamical)
$\Upsilon_V^{\rm dyn}$ lie in a fairly narrow range around 
$\widehat{\Upsilon}_V^{\rm dyn}\simeq 1.5\ M_\odot\,L_\odot^{-1}$
\citep{mcl05}. However, it is worth noting that the size of this
difference is similar to what is found in some numerical simulations
of two-body relaxation over a Hubble time in clusters with a spectrum
of stellar masses (e.g., \citealt{baumgardt03}). In such simulations,
$\Upsilon_V^{\rm dyn}$ falls below $\Upsilon_V^{\rm pop}$ due to the
preferential escape of low-mass stars with high individual
$M_*/L_*$ (population-synthesis models do not incorporate this or any
other stellar-dynamical effect). Thus, a median   
$\widehat{\Upsilon}_V^{\rm dyn} < \widehat{\Upsilon}_V^{\rm pop}$
may itself be a signature of cluster evaporation. We might then
also expect that more dynamically evolved clusters---that is, those
with shorter relaxation times---could have systematically lower
ratios of $\Upsilon_V^{\rm dyn}/\Upsilon_V^{\rm pop}$. However, this
is a relatively small effect, which is not well quantified 
theoretically and is not clearly evident in real data (the
numbers published by \citealt{mcl05} show no significant correlation
between $\Upsilon_V^{\rm dyn}/\Upsilon_V^{\rm pop}$ and $t_{\rm rh}$
for Galactic globulars). We therefore proceed, as stated, with a single
$\Upsilon_V^{\rm dyn}/\Upsilon_V^{\rm pop} = 0.8$ assumed for all GCs.

\citet{harris96} gives the projected half-light radius $R_h$ for 141
of the clusters with a mass estimated in this way, and for these we
obtain the three-dimensional half-mass radius from the general rule
$r_h=(4/3) R_h$ (\citealt{spitzer87}), which assumes no internal mass
segregation. The remaining five objects have mass estimates
but no size measurements. To each of these clusters, we assign an
$r_h$ equal to the median value for those of the other 141 GCs
having masses within a factor two of the one with unknown $r_h$.
In all cases, the half-mass density is $\rho_h \equiv 3M/8\pi r_h^3$. 

The leftmost panel in Figure \ref{fig:rhoh} shows immediately that
the cluster mass distribution has a strong dependence on half-mass
density: the median $\widehat{M}$ increases with $\rho_h$ while the
scatter in $\log\,M$---that is, the width of the GCMF---decreases. The
first of these points is related to the fact that $r_h$
correlates poorly with $M$ \citep[e.g.,][]{djorg94,mcl00}. The second
point, that the dispersion of $dN/d\,\log\,M$ decreases with
increasing $\rho_h$, is behind the finding
(\citealt{kavelaars97}; \citealt{gnedin97}) that
the GCMF is broader at very large Galactocentric radii. We return
to this in \S\ref{subsec:models}. 

A natural concern, when plotting $M$ against $\rho_h$ as we have done
here, is that any apparent correlation might only be a trivial
reflection of the definition $\rho_h \propto M/r_h^3$. This may seem
particularly worrisome because, as we just mentioned, it is known that
size does not correlate especially well with mass for GCs in the Milky
Way (or, indeed, in other galaxies). However, the lack of a tight
$M$--$r_h$ correlation does {\it not} imply that all GCs have
{\it the same} $r_h$, even within the unavoidable measurement errors.
The root-mean-square (rms) scatter of $\log\,r_h$ about its average
value is $\pm 0.3$ for Galactic GCs, and the 68-percentile spread in
$\log\,r_h$ is slightly greater than 0.5, or more than a factor of 3
in linear terms (from the data in \citealt{harris96}; see, e.g.,
Figure 8 of \citealt{mcl00}). This compares to an rms random
measurement error (from formal, $\chi^2$ fitting uncertainties) of 
$\delta(\log\,r_h) \approx 0.05$, or about 10\% relative error; and an
rms systematic measurement error (i.e., differences in the $r_h$
inferred from fitting different structural models to a single cluster)
of perhaps $\delta(\log\,r_h) \la 0.03$; see \citet{mcl05}.
Most of the scatter in plots of observed half-light radius
versus mass is therefore real and contains physical information. The
left-hand panel of Figure \ref{fig:rhoh} displays this information in
a form that highlights clear, nontrivial overall trends requiring
physical explanation.

\begin{deluxetable*}{lccccccc}
\tabletypesize{\small}
\tablecaption{Milky Way GC Properties in Bins of Density and
              Galactocentric Radius \label{tab:bins}}
\tablewidth{0pt}
\tablecolumns{8}
\tablehead{
\colhead{Bin}    & \colhead{${\cal N}$}
                 & \colhead{$\widehat{\rho}_h$\,\tablenotemark{a}}
                 & \colhead{$\widehat{r}_{\rm gc}$\,\tablenotemark{a}}
                 & \colhead{$M_{\rm min}$} & \colhead{$M_{\rm max}$}
                 & \colhead{$\widehat{M}$\,\tablenotemark{a}}
                 & \colhead{$M_{\rm TO}$\,\tablenotemark{b}}   \\
\colhead{}       & \colhead{}
                 & \colhead{[$M_\odot\,{\rm pc}^{-3}$]}  
                 & \colhead{[kpc]}                       
                 & \colhead{[$M_\odot$]} & \colhead{[$M_\odot$]}
                 & \colhead{[$M_\odot$]}                 
                 & \colhead{[$M_\odot$]}}
\startdata
\multicolumn{8}{c}{$\rho_h$ \ bins} \\
\noalign{\vskip .8ex} \hline \noalign{\vskip .8ex}
$0.034\le \rho_h \le 76.5\ M_\odot\,{\rm pc}^{-3}$
       &  48  & 8.48 & 12.9
       & $5.63\times10^2$ & $8.84\times10^5$
       & $4.12\times10^4$
       & $3.98\times10^4$ \\
$78.8\le \rho_h \le 526\ M_\odot\,{\rm pc}^{-3}$
       &  49  & 232  & 5.6
       & $8.37\times10^3$ & $1.67\times10^6$
       & $1.22\times10^5$
       & $1.58\times10^5$ \\
$579\le \rho_h\le 5.65\times10^4\ M_\odot\,{\rm pc}^{-3}$
       &  49  & 973 & 3.2
       & $1.93\times10^4$ & $1.30\times10^6$
       & $2.82\times10^5$
       & $2.88\times10^5$ \\
\noalign{\vskip .8ex} \hline \noalign{\vskip .8ex}
\multicolumn{8}{c}{$r_{\rm gc}$ \ bins} \\
\noalign{\vskip .8ex} \hline \noalign{\vskip .8ex}
$0.6\le r_{\rm gc} \le 3.2$ kpc
       &  47  & 597  & 1.9
       & $4.47\times10^3$ & $1.02\times10^6$
       & $1.15\times10^5$
       & $2.14\times10^5$ \\
$3.3\le r_{\rm gc} \le 9.4$ kpc
       &  50  & 261  & 5.2
       & $2.02\times10^3$ & $1.67\times10^6$
       & $1.27\times10^5$
       & $1.66\times10^5$ \\
$9.6\le r_{\rm gc} \le 123$ kpc
       &  49  & 18.4  & 18.3
       & $5.63\times10^2$ & $1.30\times10^6$
       & $7.42\times10^4$
       & $8.71\times10^4$
\enddata

\tablenotetext{a}{The notation $\widehat{x}$ represents the median of
                  quantity $x$.}
\tablenotetext{b}{$M_{\rm TO}$ is the peak mass of the {\it model} GCMFs
                  traced by the solid curves in each panel of Figure
                  \ref{fig:gclf}, which are given by equation
                  (\ref{eq:compschec}) of the text with $\beta=2$,
                  $M_c=10^6\ M_\odot$, and individual $\Delta$ 
                  given by the observed $\rho_h$ of each cluster
                  through equation (\ref{eq:delta}).}

\end{deluxetable*}

The dashed line in the plot of mass against density traces
the proportionality $M\propto \rho_h^{1/2}$, or 
$M r_h^3 = {\rm constant}$. Insofar as the half-mass
relaxation time scales as $t_{\rm rh} \propto (M r_h^3)^{1/2}$, and to
the extent that $\mu_{\rm ev} \propto M/t_{\rm rh} \propto \rho_h^{1/2}$ 
approximates the average rate of relaxation-driven mass loss, this
line is one of equal evaporation time. That such a locus nicely bounds
the lower envelope of the observed cluster distribution is itself a
strong hint that relaxation-driven cluster disruption has
significantly modified the GCMF at low masses (recall that
$M r_h^3 = {\rm constant}$ defines one side of the GC ``survival
triangle'' when the $M$--$\rho_h$ plot is recast as $r_h$ versus $M$:
\citealt{fall77,okazaki95,ostriker97,gnedost97}). It is also
further evidence that the weak correlation of observed $r_h$ with $M$
is due to significant and real differences in cluster radii, since if
$r_h$ were intrinsically the same for all GCs, then we would see
$M \propto \rho_h$ instead. 

The middle panel of Figure \ref{fig:rhoh} shows the well-known result
that the typical GC mass depends weakly if at all on Galactocentric  
radius, at least until large $r_{\rm gc} \ga 30$--40~kpc, where there
are too few clusters to discern any trend. The right-hand panel
of the figure shows why this is true even though the GCMF depends
significantly on cluster density: although there is a correlation
between half-mass density and Galactocentric position, the large
scatter about it is such that convolving the observed $M$ versus
$\rho_h$ with the observed $\rho_h$ versus $r_{\rm gc}$ results in an
almost null dependence of $M$ on $r_{\rm gc}$. 

We now divide the GC sample in Figure \ref{fig:rhoh} roughly into
thirds, in two different ways: first on the basis of half-mass
density, and second by Galactocentric radius. These $\rho_h$ and
$r_{\rm gc}$ bins are defined in Table \ref{tab:bins}, which also
gives a few summary statistics for the globulars in each subsample. We
count the clusters in every subsample in about 10 equal-width bins of
$\log\,M$ to obtain histogram representations of 
$dN/d\,\log\,M$, first as a function of $\rho_h$ and then as a function
of $r_{\rm gc}$. These GCMFs are shown by the points in Figure
\ref{fig:gclf}, with errorbars indicating standard Poisson
uncertainties. The curves in the figure trace model GCMFs, which
we describe in \S\ref{subsec:models}. For the moment, it is important 
to note that the dashed curve is the same in every panel, apart from
minor differences in normalization, and is proportional to the
GCMF for the whole sample of 146 GCs. (In the middle-left panel of
Figure \ref{fig:gclf}, which pertains to clusters distributed tightly
around the median $\rho_h$ of the entire GC system, the dashed curve
is coincident with the solid curve running through the data.) 

The left-hand panels of Figure \ref{fig:gclf} show directly that the
GCMF is peaked for clusters at any density, 
and that the mass of the peak increases systematically with $\rho_h$ 
(see also the last column of Table \ref{tab:bins}, but note that
the turnover masses there refer to the {\it model} GCMFs that we
develop below). The statistical significance of this is very
high, and qualitatively it is the behavior expected if
$M_{\rm TO}$ owes its existence to cluster disruption
at a rate that increases with $\rho_h$, as is the case with
relaxation-driven evaporation.

The right-hand panels of Figure \ref{fig:gclf} confirm once again
that the GCMF peak mass is a very weak function of
Galactocentric position. In fact, the observed distributions in the
two $r_{\rm gc}$ bins inside $\simeq\! 10$~kpc are statistically
indistinguishable in their entirety, and the main difference at larger
$r_{\rm gc} \ga 10$~kpc is a slightly higher proportion of low-mass
clusters rather than a large change in $M_{\rm TO}$. All of
this is consistent with the primary dependence of the GCMF being that
on $\rho_h$, since Figure \ref{fig:rhoh} shows that the GC density
distribution is not sensitive to Galactocentric position for
$r_{\rm gc}\la 10$--20~kpc but has a substantial low-density tail
at larger radii (with a broader associated GCMF, as seen in the
upper-left panel of Figure \ref{fig:gclf}).

\subsection{Simple Models}
\label{subsec:models}

\begin{figure*}
\centerline{\hfil
   \rotatebox{270}{\includegraphics[height=175mm]{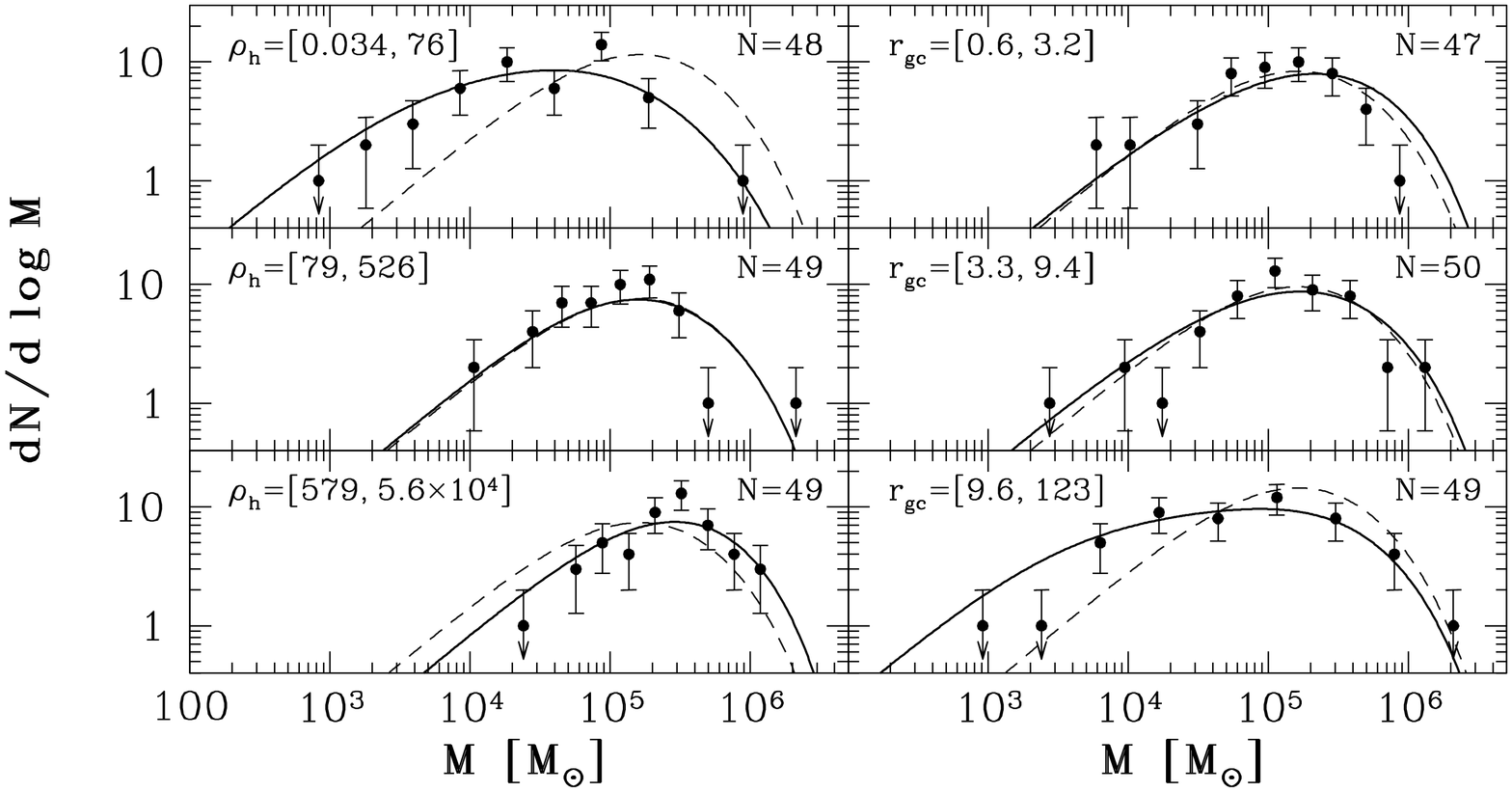}}
\hfil}
\caption{
  GCMF as a function of half-mass density,
  $\rho_h\equiv 3M/8\pi r_h^3$ ({\it left panels}), and as a function of
  Galactocentric radius, $r_{\rm gc}$ ({\it right panels}), for 146
  Milky Way GCs in the catalogue of \citet{harris96}. The dashed curve
  in all cases is an evolved \citeauthor{schechter76} function for the
  entire GC system 
  \citep{jordan07}: equation (\ref{eq:compschec}) with $\beta=2$,
  $M_c=10^6\ M_\odot$, and $\Delta \equiv 2.3\times10^5\ M_\odot$
  for all clusters (from equation [\ref{eq:delta}] and a median
  $\widehat{\rho}_h=246\ M_\odot\,{\rm pc}^{-3}$), giving a peak at 
  $M_{\rm TO}=1.6\times10^5\ M_\odot$. Solid curves are the GCMFs
  predicted by equation (\ref{eq:compschec})
  with $\beta=2$ and $M_c=10^6\ M_\odot$ but individual $\Delta$
  given by the observed $\rho_h$ of each cluster (equation
  [\ref{eq:delta}]) in the different subsamples.
\label{fig:gclf}}
\end{figure*}

We now assess more quantitatively whether these results
are consistent with evaporation-dominated evolution of the
GCMF from an initial distribution like that observed for young
clusters in the local universe. We model the time-evolution of the
distribution of $M$ versus $\rho_h$ in Figure \ref{fig:rhoh} but do
not attempt this for the distribution of $\rho_h$ over
$r_{\rm gc}$---the details of which likely depend on a complicated
interplay between the tidal field of the Galaxy, the present and past
orbital parameters of clusters, and the structural nonhomology of
GCs. To compare our models to the current GCMF as a function of
$r_{\rm gc}$, we simply calculate them using the observed $\rho_h$ of
individual clusters in different ranges of Galactocentric radius.

We assume that the initial
GCMF was independent of cluster density, and that all globulars
surviving to the present day have been losing mass for the past Hubble
time at constant rates. We use the {\it current} half-mass
density of each cluster to {\it estimate} $\mu_{\rm ev} \propto 
\rho_h^{1/2}$. As we discussed earlier, an approximately time-independent
$\mu_{\rm ev}$ is indicated by most calculations
of two-body relaxation in tidally limited GCs. We give a more
detailed, a posteriori justification in \S\ref{sec:discussion} for
using $\rho_h$, rather than other plausible measures of cluster
density, to estimate $\mu_{\rm ev}$.

Consider first a group of coeval GCs with an initial mass function 
$dN/d\,\log\,M_0$ and a single, time-independent mass-loss rate
$\mu_{\rm ev}$. The mass of every cluster decreases
linearly as $M(t) = M_0 - \mu_{\rm ev} t$, and at any later time
each has lost the same amount
$\Delta \equiv M_0-M(t) = \mu_{\rm ev} t$. \citetalias{fall01} show
rigorously that in this case, the evolved and initial GCMFs are
related by 
\begin{equation}
\frac{dN}{d\,\log\,M}
     \, = \, \frac{M}{M_0} \times \frac{dN}{d\,\log\,M_0}
     \, = \, \frac{M}{(M+\Delta)}
          \, \frac{dN}{d\,\log\,\left(M+\Delta\right)}
     \ .
\label{eq:evolve}
\end{equation}
This is the basis for the claim that the mass function scales
generically as $dN/d\,\log\,M \propto M^{+1}$ (a $\beta = 0$ power
law) at low enough $M(t) < \Delta$---that is, for
the surviving remnants of clusters with $M_0 \approx
\Delta$---just so long as the initial distribution was not a delta
function.

We follow \citetalias{fall01} (see also \citealt{jordan07}) in
adopting a \citet{schechter76} function for the initial GCMF:
\begin{equation}
dN/d\,\log\,M_0 \, \propto \,
       M_0^{1-\beta}\ \exp\,\left(-M_0/M_c\right) \ .
\label{eq:init}
\end{equation}
With $\beta\simeq 2$, this distribution describes the power-law mass
functions of young massive clusters in systems like the Antennae
galaxies \citep[e.g.,][]{zhang99}. An exponential cut-off at
$M_c \sim 10^6\ M_\odot$ is generally consistent with such data, even
if not always demanded by them; here we
require it mainly to match the curvature observed at high
masses in old GCMFs \citep[e.g.,][]{burkert00,jordan07}.

Combining equations (\ref{eq:evolve}) and (\ref{eq:init}) gives the
probability density that a single GC with known evaporation rate
and age has an instantaneous mass $M$. The
time-dependent GCMF of a system of ${\cal N}$ GCs with a range of
$\mu_{\rm ev}$ (or ages, or both) is then just the sum of all such
individual probability densities:
\begin{equation}
  \frac{dN}{d\,\log\,M} = \sum_{i=1}^{{\cal N}}
     \frac{A_i\,M}{\left[M+\Delta_i\right]^\beta} \,\,
     \exp\left[- \, \frac{M+\Delta_i}{M_c}\right] \ .
\label{eq:compschec}
\end{equation}
Here the total mass losses $\Delta_i = (\mu_{\rm ev} t)_i$ may differ
from cluster to cluster ($t_i$ being the age of a single GC) but both
$\beta$ and $M_c$ are assumed to be constants, independent of $\rho_h$
in particular.\footnotemark  
\footnotetext{Note that $M_c$ appears to take on different values in
  the GCMFs of other galaxies, varying systematically with the 
  total luminosity $L_{\rm gal}$ \citep{jordan07}. The
  reasons for this are unclear, as is the origin of this
  mass scale in the first place.}
Given each $\Delta_i$, the normalizations $A_i$ in equation
(\ref{eq:compschec}) are defined so that the integral over
$d\,\log\,M$ of each term in the summation is unity.

\citet{jordan07} have introduced a specialization of equation
(\ref{eq:compschec}) in which all clusters have the same $\Delta$. They
refer to this as an evolved \citeauthor{schechter76} function and
describe its properties in detail (including giving a formula for the
turnover mass $M_{\rm TO}$ as a function of $\Delta$ and $M_c$) for
the case $\beta=2$.
Here we note only that, at very young cluster ages or for slow
mass-loss rates, such that $\Delta \ll M_c$ and only the low-mass,
power-law part of the initial GCMF is significantly eroded, any
one evolved \citeauthor{schechter76} function has a peak at
$M_{\rm TO} \simeq \Delta/(\beta - 1)$ (for $\beta > 1$). As $\Delta$
increases relative to $M_c$, the turnover at first increases
proportionately and the width of the distribution decreases (since the
high-mass end at $M \ga M_{\rm TO}$ is largely unchanged). For large
$\Delta \gg M_c$, 
however, the peak is bounded above by $M_{\rm TO} \rightarrow M_c$ and
the width approaches a lower limit.\footnotemark
\footnotetext{The increase of $M_{\rm TO}$ and the decrease of the
  full width of $dN/d\,\log\,M$ for increasing $\Delta$ eventually
  saturate when the mass loss per GC is so high that it affects
  clusters in the exponential part of the initial
  \citeauthor{schechter76}-function GCMF. This is because
  $dN/d\,\log\,M \propto M^{+1}\,\exp(-M/M_c)$ is a self-similar
  solution to equation (\ref{eq:evolve}).} 
Thus, the dependence of $M_{\rm TO}$ on $\Delta$ is weaker
than linear when $M_c$ is finite in the initial GCMF of
equation (\ref{eq:init}). Any peak in the full equation 
(\ref{eq:compschec}) for a system of GCs with individual $\Delta$
values is an average of ${\cal N}$ different turnovers and must be
calculated numerically.

In their modeling of the Milky Way GC system, \citetalias{fall01}
effectively compute mass functions of the type
(\ref{eq:compschec})---based on the same initial conditions and 
dynamical evolution---with a distribution of $\Delta$ values
determined by the 
orbital parameters of clusters in an idealized, spherical and static
logarithmic Galaxy potential (used both to fix $\mu_{\rm ev}$ in terms
of cluster tidal densities and to estimate additional mass loss due
to gravitational shocks). \citet{jordan07} fit GCMF data in the Milky
Way and scores of Virgo Cluster galaxies with their version of
equation (\ref{eq:compschec}) in which all GCs have the same
$\Delta$. They thus estimate the dynamical mass loss from
{\it typical} clusters in these systems.
Here, we construct models for the Milky Way GCMF using $\Delta$
values given directly by the observed half-mass densities of
individual GCs.

We adopt $\beta=2$ for the initial low-mass power-law index in
equation (\ref{eq:init}), which carries over into equation
(\ref{eq:compschec}) for the evolved $dN/d\,\log\,M$.
\citet{jordan07} have fitted the full Galactic GCMF with an evolved
\citeauthor{schechter76} function assuming $\beta=2$ and a single
$\Delta \equiv \widehat{\Delta}$ for all
surviving globulars. They find $M_c \simeq 10^6\,M_\odot$ and
$\widehat{\Delta}=2.3\times10^5\,M_\odot$. We use this value of $M_c$
in equation (\ref{eq:compschec}) and we associate
$\widehat{\Delta}$ with the mass loss from clusters at the median
half-mass density of the entire GC system, which is
$\widehat{\rho}_h = 246\ M_\odot\,{\rm pc}^{-3}$ from the data in
Figure \ref{fig:rhoh}. Since we are assuming that
$\Delta = \mu_{\rm ev}t \propto \rho_h^{1/2} t$ for coeval GCs, we
therefore stipulate 
\begin{equation}
  \Delta =
     1.45\times10^4\ M_\odot\ 
     \left(\rho_h/M_\odot\,{\rm pc}^{-3}\right)^{1/2}
\label{eq:delta}
\end{equation}
for globulars with arbitrary $\rho_h$. Assuming a typical GC age of
$t=13$~Gyr, this corresponds to a mass-loss rate of
\begin{equation}
\mu_{\rm ev} \simeq
     1100\ M_\odot\,{\rm Gyr}^{-1}\
        \left(\rho_h/M_\odot\,{\rm pc}^{-3}\right)^{1/2} \ .
\label{eq:muev}
\end{equation}
In \S\ref{sec:discussion} we discuss the cluster lifetimes
implied by this value of $\mu_{\rm ev}$. We emphasize here that the
scaling of $\mu_{\rm ev}$ and $\Delta$ with $\rho_h^{1/2}$ follows
rather generically from our hypothesis of evaporation-dominated
cluster evolution, while the numerical coefficients in equations
(\ref{eq:delta}) and (\ref{eq:muev}) are specific to the assumption
of $\beta = 2$ for the power-law index at low masses in the initial
GCMF.

The dashed curve shown in every panel of Figure \ref{fig:gclf} is the
evolved \citeauthor{schechter76} function fitted to the entire GCMF of
the Milky Way by \citet{jordan07}. This has a peak at
$M_{\rm TO} \simeq 1.6\times10^5\,M_\odot$
(magnitude $M_V \simeq -7.4$ for a typical $V$-band mass-to-light
ratio of 1.5 in solar units) and gives a very good description
of the observed $dN/d\,\log\,M$ in the middle density bin,
$79\la \rho_h \la 530\ M_\odot\,{\rm pc}^{-3}$, and 
in the two inner radius bins, $r_{\rm gc} \le 9.4$~kpc. This is
expected, since the median half-mass density in each of these cluster
subsamples is very close to the system-wide median $\widehat{\rho}_h =
246\ M_\odot\,{\rm pc}^{-3}$ (see Table \ref{tab:bins}).
Even in the outermost $r_{\rm gc}$ bin, a Kolmogorov-Smirnov (KS)
test only marginally rejects the dashed-line model (at the $\simeq\! 95\%$
level), because this subsample still includes many GCs at or near the global
median $\widehat{\rho}_h$ (see Figure \ref{fig:rhoh}). By contrast,
the average GCMF is strongly rejected as a model for the lowest- and
highest-density GCs on the left-hand side of Figure \ref{fig:gclf}:
the KS probabilities that these data are drawn from the
dashed distribution are $<\! 10^{-4}$ in both cases. This
is also expected since, by construction, these bins only contain
clusters with densities well away from the median of the full GC
system, for which the total mass lost by evaporation should be
significantly different from the typical
$\widehat{\Delta} = \Delta(\widehat{\rho}_h)$.

The solid curves in Figure \ref{fig:gclf}, which are different in
every panel, are the superpositions of many different evolved
\citeauthor{schechter76} functions, as in equation
(\ref{eq:compschec}), with distinct $\Delta$ 
values given by equation (\ref{eq:delta}) using the observed $\rho_h$
of each cluster in the corresponding subsample. These models provide
excellent matches to the observed $dN/d\,\log\,M$ in every $\rho_h$
and $r_{\rm gc}$ bin, with $\chi^2<1.3$ per degree of freedom in all
cases. This is the main result of this paper.

The last column of Table \ref{tab:bins} gives the mass $M_{\rm TO}$ at
which each of the solid {\it model} GCMFs in Figure
\ref{fig:gclf} peaks. We note that these turnovers increase
roughly as $M_{\rm TO} \sim \widehat{\rho}_h^{~0.3-0.4}$ for our
specific binnings in $\rho_h$ and $r_{\rm gc}$, somewhat
shallower than the $\rho_h^{1/2}$ scaling of the cluster mass-loss
rate that defines the models. This is partly because of the
averaging over individual turnovers implied by the summation of
many evolved \citeauthor{schechter76} functions in each GC bin, and
partly because---as we discussed just after equation
(\ref{eq:compschec})---the turnover mass of any one evolved
\citeauthor{schechter76} function cannot increase
indefinitely in direct proportion to $\Delta\propto\rho_h^{1/2}$, but
has a strict upper limit of $M_{\rm TO}\le M_c$.

Our models are naturally consistent with the fact that the GCMF is
narrower for clusters with higher densities. This
is obvious in the left-hand panels of Figure \ref{fig:gclf}; in
the discussion immediately after equation (\ref{eq:compschec}), we
described how it follows from the increase of $M_{\rm TO}$ with
$\Delta \propto \rho_h^{1/2}$ for a single evolved
\citeauthor{schechter76} function. In addition,
the superposition of many such functions with separate,
density-dependent turnovers and widths results in wider
GCMFs for cluster subsamples spanning larger ranges of
$\rho_h$. This accounts in particular for the breadth of the mass
function at $r_{\rm gc}\ge 9.4$~kpc. The globulars at these radii have
$0.034\le \rho_h \le 4.1\times10^3\ M_\odot\,{\rm pc}^{-3}$,
corresponding to individual evolved \citeauthor{schechter76} functions
with turnovers at
$2.7\times 10^3 \la M_{\rm TO} \la 4.0\times10^5\ M_\odot$.
The composite GCMF in the lower-right panel of Figure \ref{fig:gclf}
is therefore extremely broad and shows a very flat peak, such that an
overall $M_{\rm TO}$ cannot be established precisely from the data
alone. This explains the findings of \citet{kavelaars97}, who pointed
out that the GCMF of the outermost third of the Milky Way cluster
system has a turnover that is statistically consistent with the
full-Galaxy average, but a larger dispersion (see also
\citealt{gnedin97}).

Finally, if the GCMF evolved dynamically from initial conditions
similar to those we have adopted, then the data and models in
the left-hand panels of Figure \ref{fig:gclf} argue against the
notion that external gravitational shocks, rather than internal
two-body relaxation, were primarily responsible for shaping the
present-day GCMF. This is because the mass-loss rate caused by
shocks alone, $-dM/dt = \mu_{\rm sh} \propto M/\rho_h$, differs
significantly from that caused by evaporation alone,
$-dM/dt = \mu_{\rm ev} \propto \rho_h^{1/2}$. The direct dependence of
$\mu_{\rm sh}$ on $M$ ensures that shocks become progressively
less important compared to evaporation as clusters lose mass
(at a given $\rho_h$),
and consequently shocks are not likely to have had much effect
on the observed GCMF for $M<M_{\rm TO}$. Furthermore, the
inverse dependence of $\mu_{\rm sh}$ on $\rho_h$ is contrary to
the direct dependence of $M_{\rm TO}$ on $\rho_h$ shown in
Figure \ref{fig:gclf}. The different roles played by shocks and
evaporation in shaping the observed GCMF are discussed more
fully by \citetalias{fall01}. We note here that gravitational
shocks may have been important in destroying very massive or
very low-density clusters early in the history of our Galaxy.

\subsection{Other Cluster Properties}
\label{subsec:conc}

If the current shape of the GCMF is fundamentally the result
of long-term cluster disruption according to a mass-loss rule like
$\mu_{\rm ev} \propto \rho_h^{1/2}$, then it should be possible to
reproduce the distribution as a function of any other cluster
attribute by using the observed $\rho_h$ of
individual GCs in equations (\ref{eq:compschec}) and (\ref{eq:delta})
to build model $dN/d\,\log\,M$ for subsamples of the
Galactic cluster system defined by that attribute---as
we did for the $r_{\rm gc}$ binning of \S\ref{subsec:models}.
Here we explore one example in which differences in
the GCMFs of two groups of globulars can be seen in this way to follow
from differences in their $\rho_h$ distributions.

\citet{smith02} have shown that
the mass function of Galactic globulars with \citet{king66} model
concentrations $c<0.99$ has a less massive peak than
that for  $c\ge 0.99$. [Here $c\equiv \log\,(r_t/r_0)$, where $r_t$
is the fitted tidal radius and $r_0$ a core scale.]
They further find that a power-law fit to the low-$c$ GCMF just below
its peak returns $dN/d\,\log\,M \propto M^{+0.5}$---shallower than the
$M^{+1}$ expected generically for a mass-loss 
rate that is constant in time---but they confirm that the
latter slope applies for the GCMF at $c\ge 0.99$. They discuss
various options to explain these results, including a suggestion that,
if the mass functions of both 
low- and high-concentration clusters evolved slowly from the same,
young-cluster--like initial distribution, then the mass-loss law for
low-$c$ GCs may have differed from that for high-$c$
clusters. However, they give no physical explanation for such a
difference, and we can show now that none is required.

\begin{figure}
\centerline{\hfil
    {\includegraphics[width=85mm]{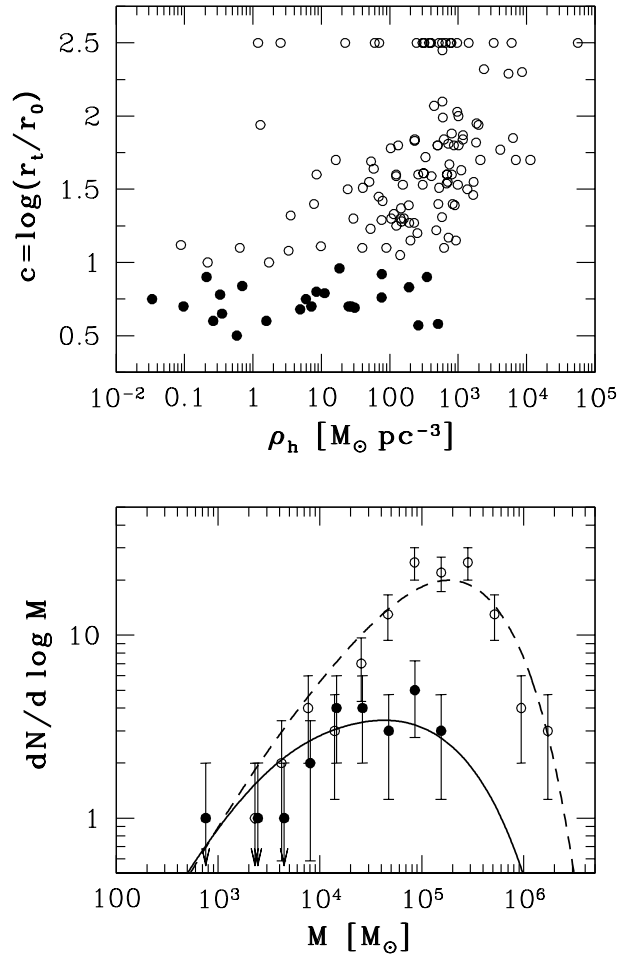}}
\hfil}
\caption{
  {\it Top:} Concentration parameter as a
  function of half-mass density for 146 Galactic GCs. The line
  of points at $c \equiv 2.5$ comes from the practice of {\it assigning}
  this value to core-collapsed clusters in the \citet{harris96} catalogue and
  its sources. {\it Bottom:} GCMF data and models
  (eqs.~[\ref{eq:compschec}] and [\ref{eq:delta}]) for 24 
  clusters with $c<0.99$ (filled circles and solid curve) and 122 clusters
  with $c\ge 0.99$ (open circles and dashed curve).
\label{fig:conc}}
\end{figure}

The upper panel of Figure \ref{fig:conc} plots 
concentration against half-mass density for the same 146 GCs from Figure
\ref{fig:rhoh};
the filled circles distinguish 24 clusters with $c<0.99$. There is a 
correlation of sorts between $c$ and $\rho_h$, 
which either derives from or causes the better-known correlation
between $c$ and $M$ \citep[e.g.,][]{djorg94,mcl00}.
The important point here is that the $\rho_h$ distribution
is offset to lower values and has a higher dispersion at $c<0.99$. 
Following the discussion in \S\ref{subsec:models}, we therefore
{\it expect} the low-concentration GCMF to have a smaller
$M_{\rm TO}$, a flatter shape around the peak, and a larger full width
than the high-concentration GCMF.

The lower panel of Figure \ref{fig:conc} shows the GCMFs for
$c<0.99$ (filled circles) and $c\ge 0.99$ (open circles). The curves are
again given by equation (\ref{eq:compschec}) with $\beta=2$,
$M_c=10^6\ M_\odot$, and individual $\Delta$ calculated from the
observed cluster $\rho_h$ 
through equation (\ref{eq:delta}). These models peak at
$M_{\rm TO}\simeq 4.3\times 10^4\ M_\odot$ for the $c<0.99$ subsample
but at $M_{\rm TO}\simeq 1.8\times10^5\ M_\odot$ for $c\ge 0.99$,
entirely as a result of the different $\rho_h$ involved. The larger width of
$dN/d\,\log\,M$ and its shallower slope at any $M\la 10^5\ M_\odot$ for the
low-concentration GCs are also clear, in the model curves as well as
the data. It is further evident that there are no low-$c$ Galactic 
globulars observed with $M\ga 2\times10^5\,M_\odot$, above the
nominal turnover of the full GCMF (as \citealt{smith02} noted). But
this is not surprising, given that there are so few low-concentration
clusters in total and they are expected to be dominated by low-mass
objects because of their generally low densities. Thus, the solid
curve in Figure \ref{fig:conc} predicts perhaps $\simeq\!3$ high-mass
clusters with $c<0.99$, where none is found.

The apparent variation of the Milky Way GCMF with
internal concentration is therefore consistent with
the same density-based model for evaporation-dominated dynamical
evolution that we compared to $dN/d\,\log\,M$ as a function of
$\rho_h$ and $r_{\rm gc}$ in \S\ref{subsec:models}. To show this, we
have made use of the densities $\rho_h$ exactly as observed within the
two concentration bins indicated in Figure \ref{fig:conc}---just as we
also took $\rho_h$ directly from the data for GCs in different ranges
of $r_{\rm gc}$ to construct models for comparison with the observed
$dN/d\,\log\,M$ in the right-hand panels of Figure \ref{fig:gclf}.
Of course, this is not the same as {\it explaining} the distribution
of $\rho_h$ versus $r_{\rm gc}$ or $c$. Doing so 
would certainly be of interest in its own right, but it is beyond the
scope of our work here.

\section{Discussion}
\label{sec:discussion}

In this section, we first show that the mass-loss rate in equation
(\ref{eq:muev}) above implies cluster lifetimes that compare
favorably with those expected from
relaxation-driven evaporation. Then we discuss why it is
reasonable to approximate $\mu_{\rm ev} \propto \rho_h^{1/2}$ in the
first place. Finally, we address the issue of possible conflict, in
some other models for evaporation-dominated GCMF evolution, between
the near-constancy of $M_{\rm TO}$ as a function of $r_{\rm gc}$ and
the observed kinematics of GC systems.

\subsection{Cluster Lifetimes}
\label{subsec:lifetime}

The disruption time of a GC with mass $M$ and a
steady mass-loss rate $\mu_{\rm ev}$ is just
$t_{\rm dis} = M/\mu_{\rm ev}$. It is convenient, for purposes of
comparison with evaporation times in the literature,
to normalize $t_{\rm dis}$ to the relaxation time of a cluster at its
half-mass radius. In general, this is 
$t_{\rm rh} = 0.138 M^{1/2} r_h^{3/2}/
                 \left[G^{1/2} m_* \ln\,(\gamma M/m_*)\right]$,
where $m_*$ is the mean stellar mass. For clusters of stars with a
single mass, $m_* \simeq 0.7\,M_\odot$ and $\gamma \simeq 0.4$ are
appropriate (\citealt{spitzer87}; \citealt{bt87}, equation [8-72]), in
which case equation (\ref{eq:muev}) for $\mu_{\rm ev}$ from our GCMF
modeling implies
\begin{equation}
  \frac{t_{\rm dis}}{t_{\rm rh}}
       \ =
       \ \frac{M}{\mu_{\rm ev} t_{\rm rh}}
       \ \simeq \ 10
       \ \left[\frac{\ln \left(0.57\, M/M_\odot\right)}
                    {\ln \left(0.57 \times 10^5\right)}\right]
       \ .
\label{eq:tdis}
\end{equation}
Clusters with realistic stellar mass spectra will have slightly
different values of $m_*$ and a smaller
$\gamma$ in the calculation of the relaxation time \citep{giersz96},
which changes the numerical value of $t_{\rm dis}/t_{\rm rh}$ somewhat
but does not alter any scalings.

We obtained the normalization of $\mu_{\rm ev} \propto \rho_h^{1/2}$
in \S\ref{subsec:models} by fitting to observed GCMFs constructed by
applying a specific mass-to-light ratio $\Upsilon_V$ to every
cluster, with models assuming a specific form for the initial
$dN/d\,\log\,M_0$. Thus, the
result in equation (\ref{eq:tdis}) depends both on the median
$\widehat{\Upsilon}_V$ and on the power-law index $\beta$ at
low masses in the original \citeauthor{schechter76}-function GCMF. The
net scaling, for either single- or multiple-mass clusters, is
\begin{equation}
t_{\rm dis}/t_{\rm rh} \propto
               \widehat{\Upsilon}_V^{-1/2} (\beta-1)^{-1} \ .
\label{eq:tdis_scale}
\end{equation}

To see the dependence of this dimensionless lifetime on
$\widehat{\Upsilon}_V$, note that we require
$\mu_{\rm ev} \propto \Delta \propto \Upsilon_V$ to fit the
mass losses of clusters with a given distribution of luminosities (the
direct observables), whereas $M/t_{\rm rh}$ is proportional to
$\rho_h^{1/2} \propto \Upsilon_V^{1/2} (L/r_h^3)^{1/2}$. Therefore,
$t_{\rm dis}/t_{\rm rh} \propto (M/t_{\rm rh})/\mu_{\rm ev}
      \propto \Upsilon_V^{-1/2}$.
The mass-to-light ratios adopted in this paper, with a median value
$\widehat{\Upsilon}_V \simeq 1.5\ M_\odot\,L_\odot^{-1}$, are
tied directly to dynamical determinations (\S\ref{subsec:data}).
 
To understand the dependence on $\beta$ in equation
(\ref{eq:tdis_scale}), recall first that the coefficients in our
expressions for $\Delta$ and $\mu_{\rm ev}$
as functions of $\rho_h$ (eqs.~[\ref{eq:delta}] and [\ref{eq:muev}])
followed from choosing $\beta=2$ for the power-law exponent 
at low masses in the initial GCMF (equation [\ref{eq:init}]).
As we mentioned just after equation (\ref{eq:compschec}), the turnover
mass of an evolved \citeauthor{schechter76} function with any
$\beta > 1$ is $M_{\rm TO}\simeq \Delta/(\beta-1)$ in the limit of low
$\Delta \propto \rho_h^{1/2}$, and $M_{\rm TO}\rightarrow M_c$ for
very high  $\Delta$. In this sense, the strongest observational
constraints on the normalizations of $\Delta$ and $\mu_{\rm ev}$ come
from the low-density clusters. All other things being equal,
their GCMF can be reproduced with $\beta \ne 2$ if $\Delta$
and $\mu_{\rm ev}$ are multiplied by $(\beta-1)$ at fixed
$\rho_h$. Therefore, 
$t_{\rm dis} \propto 1/\mu_{\rm ev} \propto 1/(\beta-1)$.
Observations of young massive clusters \citep[e.g.,][]{zhang99}
indicate that $\beta$ is near 2; but if it were slightly
shallower, then the cluster lifetimes we infer from the old GCMF would
increase accordingly. Even a relatively minor change to $\beta=1.5$
would double $t_{\rm dis}/t_{\rm rh}$ from $\approx\! 10$ to
$\approx\! 20$.

In the model of \citet{henon61} for single-mass clusters evolving
self-similarly
(fixed ratio $\rho_t/\rho_h$ of mean densities inside the
tidal and half-mass radii) in a steady tidal field ($\rho_t$ constant
in time), a cluster loses 4.5\% of its remaining mass
every half-mass relaxation time. The time to complete disruption is
therefore $t_{\rm dis}/t_{\rm rh} = 1/0.045 \simeq 22$.
For non-homologous clusters in a steady tidal
field, $t_{\rm dis}/t_{\rm rh}$ is a function of central
concentration and can differ from the \citeauthor{henon61} value by
factors of about two. From one-dimensional Fokker-Planck calculations,
\citet{gnedost97} find $t_{\rm dis}/t_{\rm rh} \simeq 10$--40 for
\citet{king66} model clusters with $c$ values similar to those
found in real GCs and with gravitational shocks suppressed (see their
Figure 6). Thus, even though the evaporation time in equation
(\ref{eq:tdis}) may be slightly shorter than is typically
found in theoretical calculations, it is certainly within the
range of such calculations. Moreover, the 
assumptions of a steady tidal field and a single stellar mass in  
\citet{henon61} and \citet{gnedost97} are important. Part of the
difference between the typical lifetimes in these particular
theoretical treatments and our estimate of
$t_{\rm dis}/t_{\rm rh}$ from the GCMF is that the former do not
include gravitational shocks, which may have accelerated somewhat the
evolution of real clusters (although we stress
again that shocks do not appear in general to have dominated the
evolution of extant Galactic GCs and are not expected to affect the
basic time-independence of the net mass-loss rate; see
\citealt{vesperini97b},
\citealt{gnedin99}, \citetalias{fall01}, and \citealt{prieto06}). A
spectrum of stellar masses in the clusters may also have
contributed to an increase in evaporation rate over the
single-mass values \citep[e.g.,][]{johnstone93,lee95}.

Estimates of evaporation times from other numerical methods and for
models of multimass clusters can be rather sensitive to the detailed
computational techniques and input assumptions and approximations, and
differences at roughly the factor-of-two level in
$t_{\rm dis}/t_{\rm rh}$ between different analyses are not uncommon;
see, e.g., 
\citet{vesperini97b}, \citet{takahashi98,takahashi00}, 
\citet{baumgardt01}, \citet{joshi01}, \citet{giersz01}, and
\citet{baumgardt03}. Thus, although the lifetimes in these studies
tend to be broadly comparable to those in \citet{henon61} and
\citet{gnedost97}, noticeably shorter values do occur in some
models. In any case, we are encouraged by consistency to within
factors of two or three between estimates of $t_{\rm dis}$ or
$\mu_{\rm ev}$ by such vastly different methods---one purely
observational, based on the mass functions of cluster systems; the
other purely theoretical, based on idealized models for the evolution
of individual clusters---particularly since each method involves
several uncertain inputs and parameters.

\subsection{Approximating $\mu_{\rm ev} \propto \rho_h^{1/2}$}
\label{subsec:approx}

\subsubsection{Half-mass versus Tidal Density}
\label{subsubsec:rhvrt}

The dimensionless disruption time in equation (\ref{eq:tdis}) is
independent of any cluster property other than the Coulomb logarithm
because we have used GC half-mass densities to estimate
$t_{\rm dis} = M/\mu_{\rm ev} \propto M/\rho_h^{1/2}$, while
$t_{\rm rh}$ also scales as 
$M/\rho_h^{1/2}$. However, as we mentioned
above, the Fokker-Planck calculations of \citet{gnedost97} in
particular show that $t_{\rm dis}/t_{\rm rh}$ is actually a 
function of central concentration, $c$, for \citet{king66} model
clusters in steady tidal fields. The constant of proportionality in
$\mu_{\rm ev} \propto \rho_h^{1/2}$ should therefore also
depend on $c$, a detail that we have neglected to this point. We show
now that this has not biased any of our analysis or affected our
conclusions.

\begin{figure}
\centerline{\hfil
    {\includegraphics[width=85mm]{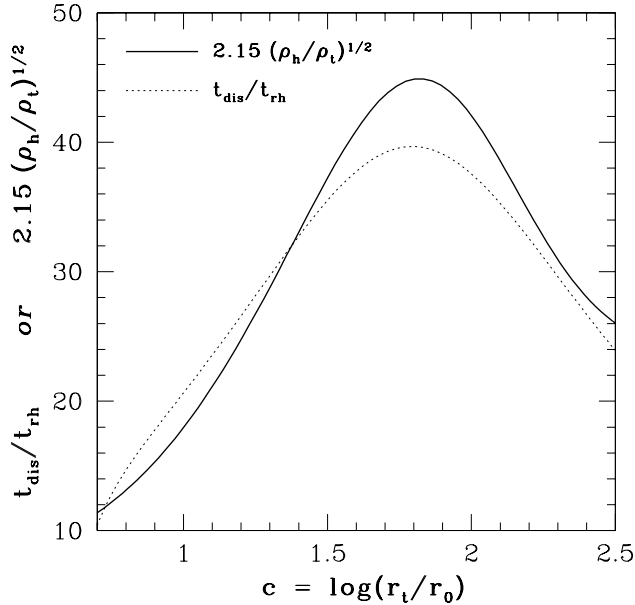}}
\hfil}
\caption{
  Dependence of $t_{\rm dis}/t_{\rm rh}$ (dotted line; from
  \citealt{gnedost97}) and $(\rho_h/\rho_t)^{1/2}$ (solid line; after
  scaling by a factor of 2.15) on central concentration for
  single-mass \citeauthor{king66}-model clusters. Over the range of
  $c$ shown, which includes nearly all Galactic globulars, the 
  approximate proportionality
  $t_{\rm dis}/t_{\rm rh} \propto (\rho_h/\rho_t)^{1/2}$ holds
  to within better than 15\%. Thus, to this level of accuracy the
  evaporation time $t_{\rm dis}$ is roughly the same multiple
  of $M/\rho_t^{1/2}$ for clusters with any internal density profile.
\label{fig:life}}
\end{figure}

The dotted curve in Figure \ref{fig:life} illustrates the dependence of
$t_{\rm dis}/t_{\rm rh}$ on $c$ for single-mass
\citeauthor{king66} models, as given by equation (30) of
\citet{gnedost97}. The solid curve is proportional to
$(\rho_h/\rho_t)^{1/2} = (r_t^3/2r_h^3)^{1/2}$, which
we have calculated as a function of $c$ for these models
and multiplied by a constant to compare directly with
$t_{\rm dis}/t_{\rm rh}$. Evidently, there is an approximate equality
$t_{\rm dis}/t_{\rm rh} \approx 2.15 (\rho_h/\rho_t)^{1/2}$, which
holds to within $<\! 15\%$ over the range of concentrations shown
in Figure \ref{fig:life} (note that all but 6 Galactic GCs have
$0.7\le c \le 2.5$, corresponding to central potentials
$3\la W_0 \la 11$). Thus, if the evaporation time is written as
$t_{\rm dis} \propto t_{\rm rh}(\rho_h/\rho_t)^{1/2}
             \propto M/\rho_t^{1/2}$,
then the constant of proportionality in the mass-loss rate
$\mu_{\rm ev} \propto M/t_{\rm dis} \propto \rho_t^{1/2}$
should be nearly independent of $c$. In fact, \citet{king66}
originally concluded, from quite basic arguments, that the
evaporation rate of a cluster with a lowered-Maxwellian velocity
distribution would take the form $\mu_{\rm ev} \propto \rho_t^{1/2}$
with only a weak dependence on $c$. An essentially
concentration-independent scaling of $\mu_{\rm ev}$ with
$\rho_t^{1/2}$ is also found in $N$-body simulations of tidally
limited, multimass clusters (e.g., \citealt{vesperini97b}) and so is
not an artifact of any assumptions specific to the
calculations of either \citet{king66} or \citet{gnedost97}.

This suggests that it might have been more natural to specify cluster
evaporation rates proportional to $\rho_t^{1/2}$ rather than
$\rho_h^{1/2}$ when developing our GCMF models in
\S\ref{sec:results}. For any cluster in a steady tidal field, with
a constant $\rho_t$, such a choice would also have been automatically
consistent with an approximately time-independent $\mu_{\rm ev}$ and
the corresponding linear $M(t)$ dependence that we have adopted
throughout this paper. As we discussed at the beginning of
\S\ref{sec:results}, our decision to work with $\rho_h$ rather than
$\rho_t$ was motivated by the fact that the half-mass density is much
better defined in principle and more accurately observed in
practice. Nevertheless, re-writing $\mu_{\rm ev} \propto \rho_t^{1/2}$
as $\mu_{\rm ev} \propto (\rho_t/\rho_h)^{1/2} \times \rho_h^{1/2}$
makes it clear that the validity of our models, with
a fixed coefficient in $\mu_{\rm ev} \propto \rho_h^{1/2}$, depends 
on the extent to which variations in $(\rho_t/\rho_h)^{1/2}$ can safely
be ignored.

Figure \ref{fig:life} shows that the full range of possible values
for $(\rho_h/\rho_t)^{1/2}$ in \citeauthor{king66}-model clusters with
$c\ge 0.7$ is only a factor of $\simeq\! 4$ between minimum and maximum.
Therefore, using a single, intermediate value of this density
ratio to describe all GCs (or a single GC evolving in time through a
series of quasi-static \citeauthor{king66} models)---which we have
effectively done by using a GCMF fit to normalize
$\Delta$ and $\mu_{\rm ev}$ in equations
(\ref{eq:delta}) and (\ref{eq:muev})---should never be in
error by more than a factor of 2 or so. This is a relatively small
inaccuracy, given that measured GC densities range over four to five
orders of magnitude.

To confirm more directly that our models with
$\mu_{\rm ev} \propto \rho_h^{1/2}$
are good approximations to GCMF evolution under a mass-loss
law $\mu_{\rm ev} \propto \rho_t^{1/2}$, we have repeated the
analysis of \S\ref{sec:results} in full but using the GC
tidal densities $\rho_t$ (derived from the values of $r_t$ listed by
\citealt{harris96}) in place of $\rho_h$  throughout. All of our main
results persist.

\begin{figure*}
\centerline{\hfil
    {\includegraphics[width=165mm]{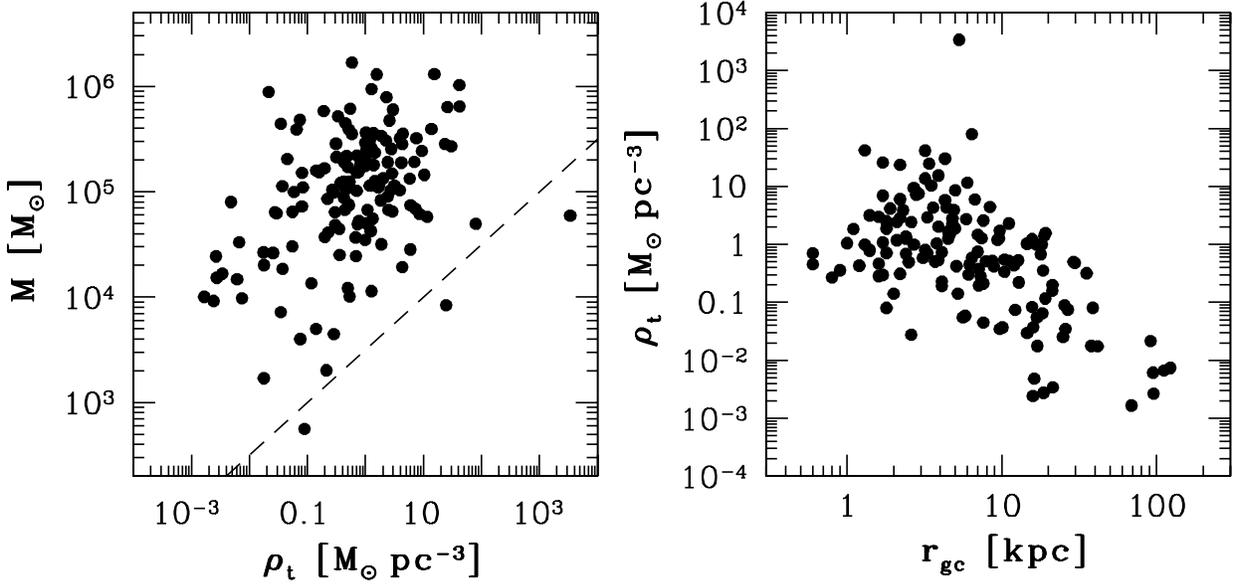}}
\hfil}
\caption{
  Scatter plots of mass $M$ versus mean density inside the tidal radius
  ($\rho_t\equiv 3M/4\pi r_t^3$) and of $\rho_t$ versus Galactocentric
  radius $r_{\rm gc}$, for 146 Galactic GCs from the \citet{harris96}
  catalogue. These plots are analogous to the left- and rightmost
  panels of Figure \ref{fig:rhoh}. The dashed line in the left-hand
  plot traces the relation $M\propto \rho_t^{1/2}$, which defines a
  locus of constant evaporation time for
  $\mu_{\rm ev} \propto \rho_t^{1/2}$.
\label{fig:rhot}}
\end{figure*}

For example,
the two panels of Figure \ref{fig:rhot}, which are analogous to the
left- and rightmost panels of Figure \ref{fig:rhoh} above, show that
(1) the GC mass distribution has a clear dependence on $\rho_t$, with
a lower envelope that is well matched by a line of constant
evaporation time, $M\propto \rho_t^{1/2}$ (the dashed line in the
plot); and (2) although the scatter in the distribution of $\rho_t$
over Galactocentric radius is smaller than the scatter in $\rho_h$
versus $r_{\rm gc}$, it is still significant.
Because the $M$--$r_{\rm gc}$ distribution can now
be viewed as the convolution of the $M$--$\rho_t$ distribution with the
$\rho_t$--$r_{\rm gc}$ distribution, the scatter in $\rho_t$ versus
$r_{\rm gc}$ is again critical in explaining the weak or null
dependence of the GCMF on Galactocentric radius. (The $M$--$r_{\rm gc}$
distribution is, of course, unchanged from that shown in the middle
panel of Figure \ref{fig:rhoh}.)\footnotemark
\footnotetext{As was also the case with our earlier plots
  involving $\rho_h$ in Figure \ref{fig:rhoh}, the scatter and
  structure in both panels of Figure \ref{fig:rhot} are real, since
  the rms scatter of $\log\,r_t$ about the best-fit lines to either of
  $\log\,M$ or $\log\,r_{\rm gc}$ is 0.3--0.35 while the
  rms errorbars based on formal fitting uncertainties are in the range
  $\delta(\log\,r_t)\simeq 0.05$--0.15 for a variety of models
  \citep{mcl05}.}  

\begin{figure*}
\centerline{\hfil
    \rotatebox{270}{\includegraphics[height=175mm]{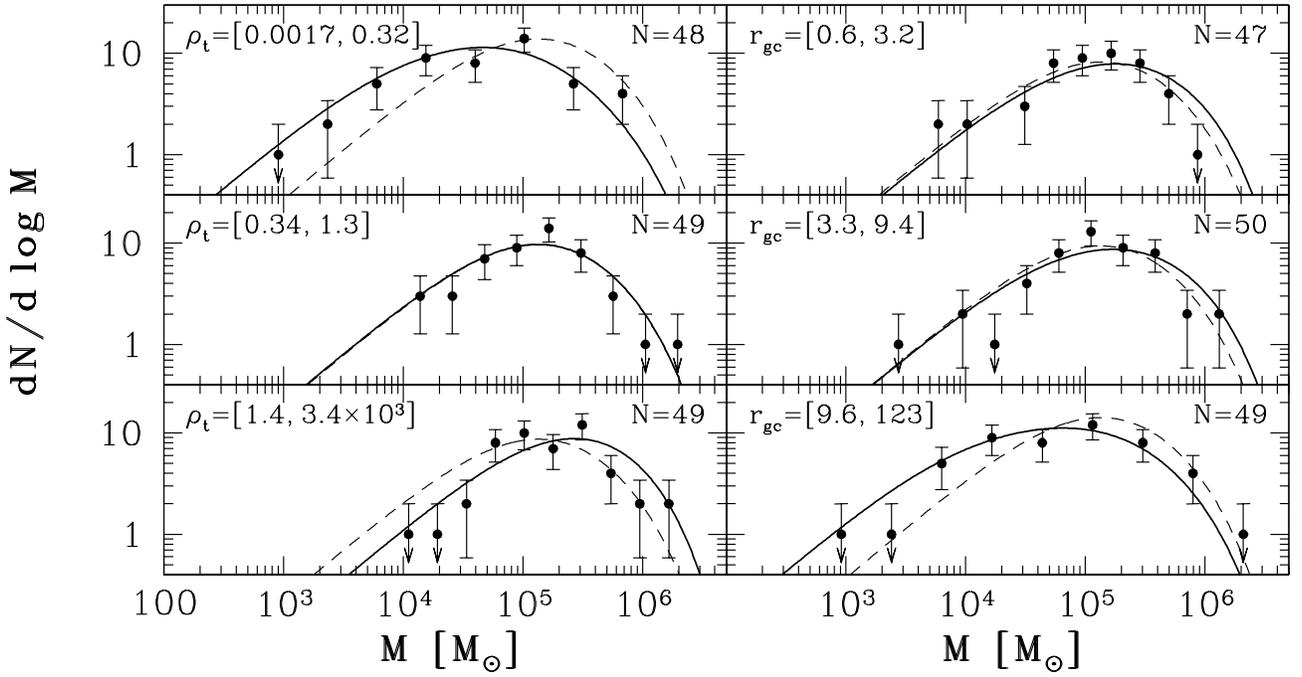}}
\hfil}
\caption{
  Observed GCMF (points, with Poisson errorbars) and models (curves)
  as a function of mean cluster density inside the tidal radius,
  $\rho_t \equiv 3M/4\pi r_t^3$ (left-hand panels), and as a
  function of Galactocentric radius, $r_{\rm gc}$ (right-hand
  panels). The dashed curve in every panel is an evolved
  \citeauthor{schechter76} function representing the entire GC system:
  equation (\ref{eq:compschec}) with $\beta=2$, $M_c=10^6\,M_\odot$,
  and a single $\Delta$, common to all clusters, evaluated from
  equation (\ref{eq:deltat}) using the median $\widehat{\rho}_t$ of
  all 146 Galactic GCs. Solid curves are subsample-specific models
  using equation (\ref{eq:compschec}) with $\beta=2$ and
  $M_c=10^6\,M_\odot$ but a different $\Delta$ value for every cluster
  (obtained from equation [\ref{eq:deltat}] using individual
  observational estimates of $\rho_t$) in any $\rho_t$ or $r_{\rm gc}$
  bin. 
\label{fig:rhot_gclf}}
\end{figure*}

Figure \ref{fig:rhot_gclf} shows the Milky Way GCMF for globulars in
three equally populated bins of tidal density (defined as indicated in
the left-hand panels of the plot) and in the same three bins of
Galactocentric radius that we used in \S\ref{subsec:models} above.
Our models for these distributions are based as before on
equation (\ref{eq:compschec}) with $\beta=2$, but now the total mass
lost from any GC is estimated from its tidal density rather than its
half-mass density. Specifically, we take
\begin{equation}
  \Delta =
     2.1\times10^5\ M_\odot\ 
     \left(\rho_t/M_\odot\,{\rm pc}^{-3}\right)^{1/2} \ .
\label{eq:deltat}
\end{equation}
The numerical coefficient in equation (\ref{eq:deltat}) is such that
it gives a $\Delta$ identical to that in equation (\ref{eq:delta})
for a GC with $\rho_h/\rho_t = 210$, which is
the median value of this density ratio for the 146 GCs in the
\citet{harris96} catalogue.

As in Figure \ref{fig:gclf}, the dashed curve in every panel of
Figure \ref{fig:rhot_gclf} is the same, representing a fit to the
average $dN/d\,\log\,M$ of the entire Galactic GC system. Thus, it is
immediately clear that the peak mass of the GCMF increases
significantly and systematically with increasing $\rho_t$, just as it
does with increasing $\rho_h$. Meanwhile, the solid curves are
subsample-specific model GCMFs, obtained by using the observed tidal
density of each cluster in any $\rho_t$ or $r_{\rm gc}$ bin to specify
individual $\Delta$ values via equation (\ref{eq:deltat}) for each of
the evolved \citeauthor{schechter76} functions in the summation of
equation (\ref{eq:compschec}).   
As expected, there is no appreciable difference, in terms of the fits
to any of the observed GCMFs, between these models based on
evaporation rates $\mu_{\rm ev} \propto \rho_t^{1/2}$ and our original
models with $\mu_{\rm ev} \propto \rho_h^{1/2}$.

\subsubsection{Retarded Evaporation}
\label{subsubsec:nonstand}

Another potential concern comes from recent arguments (see especially
\citealt{baumgardt01,baumgardt03}) that the total evaporation
time of a tidally limited cluster is not simply a multiple of an
internal two-body relaxation time, 
$t_{\rm rlx} \propto (M r^3)^{1/2}$, but depends on both
$t_{\rm rlx}$ and the crossing time
$t_{\rm cr} \propto (M/r^3)^{-1/2}$ through the combination
$t_{\rm dis} \propto t_{\rm rlx}^x t_{\rm cr}^{1-x}$,
with $x < 1$. The mass-loss rate $\mu_{\rm ev} \propto M/t_{\rm dis}$
then scales as  $M^{3/2 - x} r^{-3/2}$, which for 
$x\ne 1$ differs from the rates $\mu_{\rm ev} \propto \rho_h^{1/2}$
and $\mu_{\rm ev} \propto \rho_t^{1/2}$ that we have so far adopted.
However, our GCMF models are still meaningful, because
postulating $t_{\rm dis} \propto t_{\rm rlx}^x t_{\rm cr}^{1-x}$ 
implies a dependence of $\mu_{\rm ev}$ on a measure of cluster density
that is, once again, well {\it approximated} by $\rho_h^{1/2}$
for Galactic GCs. Before showing this, we briefly discuss the reasons
and the evidence for a possible dependence of $t_{\rm dis}$
on both $t_{\rm rlx}$ and $t_{\rm cr}$. 

If stars are assumed to escape a cluster as soon as they have attained
energies above some critical value as a result of two-body
relaxation, then $t_{\rm dis} \propto t_{\rm rlx}$ is expected (and
confirmed by $N$-body simulations; e.g., \citealt{baumgardt01}).
However, more complicated behavior may arise when escape not only
depends on stars satisfying such an energy criterion, but also
requires them to cross a spatial boundary. Then, although the stars
are still scattered to near- and above-escape energies on the
timescale $t_{\rm rlx}$, they require some additional time to actually
leave the cluster. This escape timescale is related fundamentally to
$t_{\rm cr}$ (but also depends on details of the stellar
orbits, the external tidal field, and the shape of the zero-energy
surface). The longer this extra time, the higher is the probability
that further encounters with bound cluster stars may scatter any
potential escapers back down to sub-escape energies. The net result is
a slow-down (``retardation'') of the overall evaporation rate
\citep{chandra42,king59,takahashi98,takahashi00,fukushige00,baumgardt01}
and a lengthening of the cluster lifetime $t_{\rm dis}$,
by a factor that can be expected to increase with the ratio
$t_{\rm cr}/t_{\rm rlx}$. If this factor scales as
$(t_{\rm cr}/t_{\rm rlx})^{1-x}$ for some $x<1$, then
$t_{\rm dis} \propto t_{\rm rlx}\, (t_{\rm cr}/t_{\rm rlx})^{1-x}
             = t_{\rm rlx}^{x} t_{\rm cr}^{1-x}$.

While such a retardation of evaporation can be expected to occur at some
level in all clusters, there are physical subtleties in the effect
that are probably not captured adequately by a simple
re-parametrization of lifetimes as
$t_{\rm dis} \propto t_{\rm rlx}^x t_{\rm cr}^{1-x}$.
In particular, it is unlikely that this expression
can hold for clusters of all masses with a single value
of $x<1$. Since $t_{\rm cr}/t_{\rm rlx} \propto M^{-1}$, very massive
clusters have $t_{\rm cr} \ll t_{\rm rlx}$, and stars scattered to
greater than escape energies by relaxation cross the tidal boundary
effectively instantaneously---implying that the standard
$t_{\rm dis} \propto t_{\rm rlx}$, or $x\rightarrow 1$, applies in the
high-mass limit. Indeed, if this were not the case, and a fixed $x<1$
held for all $M$, then an unphysical $t_{\rm dis}<t_{\rm rlx}$ would
obtain at high enough masses; see \citet{baumgardt01} for further
discussion. Unfortunately, ``very massive'' is not well quantified in
this context, and it is not yet clear if a single value of $x$ is
accurate for the entire GC mass regime. So far, it has been
checked directly only for initial cluster masses below the current
peak of the GCMF.

It is also worth noting that the analysis and simulations aimed at
this problem to date have dealt with clusters on circular or
moderately eccentric orbits in galactic potentials that are static and
spherical. This means that any tidal perturbations felt by stars
within the clusters are relatively weak and/or slow compared to their
own orbital periods, leading to nearly adiabatic or at least
non-impulsive responses. In more realistic situations, the galactic
potential would be time-dependent and non-spherical and there might be
additional tidal perturbations, including disk and bulge shocks. These
perturbations could in some cases {\it accelerate} the escape of 
weakly bound stars from the clusters and thus counteract the
retardation effect to some degree.
Further study is therefore needed to determine the regime of validity
of the formula $t_{\rm dis} \propto t_{\rm rlx}^x t_{\rm cr}^{1-x}$
and its possible modification outside this regime.

In the meantime,
\citet{baumgardt01} and \citeauthor{baumgardt03}
(\citeyear{baumgardt03}; hereafter \citetalias{baumgardt03}) have 
fitted this formula to the lifetimes of a suite of $N$-body
clusters with initial masses $M_0 \la 7\times 10^4\,M_\odot$ and
several different initial concentrations and orbital eccentricities.
\citetalias{baumgardt03} at first write $t_{\rm dis}$ in terms of
the relaxation and crossing times of clusters at their half-mass
radii, so that $t_{\rm rlx} \propto (M r_h^3)^{1/2}$, 
$t_{\rm cr} \propto (M/r_h^3)^{-1/2}$, and
$t_{\rm dis} \propto M^{x-1/2} r_h^{3/2}$ (see their equation
[5]). However, they immediately take a factor of
$(r_t/r_h)^{3/2}$ out from the normalization of this scaling---in
effect to obtain 
$t_{\rm dis} \propto M^{x-1/2} r_t^{3/2}$ with a different constant of
proportionality---and then use a simple definition of the tidal radius
(their equation [1], $r_t^3 = G M r_p^2/2 V_c^2$, which is appropriate
for a circular orbit of radius $r_p$ in a logarithmic potential with
circular speed $V_c$; see \citealt{innanen83}) to obtain the total
lifetime of a cluster as a 
function of its initial mass, perigalactic distance, and
$V_c$ (their equation [7]). A single exponent $x \simeq 0.75$ and a
single normalization in this function then suffice to predict
to within 10\% the lifetimes of the simulated clusters,
regardless of their initial
concentrations. By implication, if $t_{\rm rlx}$ and $t_{\rm cr}$
were fixed at $r_h$ rather than $r_t$, then $t_{\rm dis}$
{\it would} have an additional concentration dependence, related to
the ratio $(r_t/r_h)^{3/2}$---very similar to what we discussed in
\S\ref{subsubsec:rhvrt} for the case $x=1$.

We now re-examine the Milky Way GCMF in terms of this prescription for
retarded evaporation (bearing in mind the caveats mentioned above).
To avoid any explicit dependences on concentration, we
also focus on the tidal radius and write
$t_{\rm dis} \propto M^{x-1/2} r_t^{3/2}$ for general $x\le 1$; but we
do not substitute a potential- and orbit-specific formula for $r_t$ in
terms of $r_p$ and galactic properties such as $V_c$. Instead, to keep
the emphasis entirely on cluster densities, we re-write the scaling of
the lifetime in terms of the mean {\it surface} density inside the
tidal radius, $\Sigma_t \equiv M/\pi r_t^2$, and the corresponding
volume density $\rho_t = 3M/4\pi r_t^3$. This leads to
$t_{\rm dis} \propto M\, \Sigma_t^{-3(1-x)} \rho_t^{-2(x-3/4)}$,
which then implies
\begin{equation}
\mu_{\rm ev} \ \equiv\ -dM/dt
  \ \propto\ M/t_{\rm dis}
  \ \propto\ \Sigma_t^{3(1-x)} \rho_t^{2(x-3/4)} \ .
\label{eq:muev_x}
\end{equation}
Clearly, the standard $\mu_{\rm ev} \propto \rho_t^{1/2}$, which we
have already discussed, is recovered for $x=1$; while for
$x = 0.75$, we have the equally straightforward
$\mu_{\rm ev} \propto \Sigma_t^{3/4}$. 

\citetalias{baumgardt03} find that, even with the retarded evaporation
implied by $x \simeq 0.75$, the masses of their simulated clusters
still decrease approximately linearly with time after
stellar-evolution effects (which are only important for the first few
$10^8$~yr) are separated out; see especially their Figure 6,
equation (12), and related discussion. Thus, if the GCMF initially
rose towards low masses and has been eroded by slow, relaxation-driven
cluster destruction, then in this modified description of evaporation
we might expect the current mass function to depend fundamentally on
$\Sigma_t$ rather than $\rho_h$ or $\rho_t$. But because $M(t)$ still
decreases nearly linearly with $t$, only now with
$\mu_{\rm ev} \propto \Sigma_t^{3/4}$ for each cluster, the shape of
the evolved GCMF and its dependence on $\Sigma_t$ should resemble our
earlier results for $\rho_h$ and $\rho_t$.

We have confirmed this expectation by repeating all of our analyses in
\S\ref{sec:results} again, now using $\mu_{\rm ev} \propto
\Sigma_t^{3/4}$ to estimate cluster mass-loss rates. As before, we
calculate $\Sigma_t$ from the data in the \citet{harris96} catalogue,
although we caution once more that the tidal radii, and
thus the derived $\Sigma_t$, are more uncertain than $r_h$ and
$\rho_h$.

\begin{figure*}
\centerline{\hfil
    {\includegraphics[width=165mm]{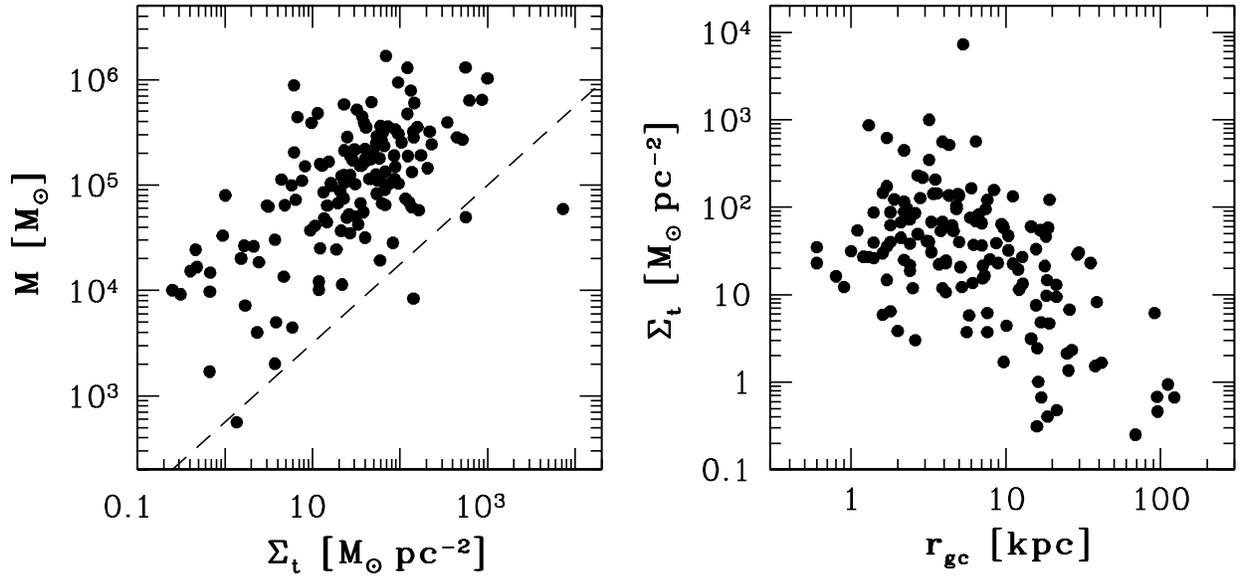}}
\hfil}
\caption{
  Scatter plots of mass $M$ versus mean {\it surface} density
  inside the tidal radius ($\Sigma_t\equiv M/\pi r_t^2$) and of
  $\Sigma_t$ versus Galactocentric radius $r_{\rm gc}$, for 146
  Galactic GCs from the \citet{harris96} catalogue. These plots are
  analogous to the left- and rightmost panels of Figure
  \ref{fig:rhoh}, and the two panels of Figure \ref{fig:rhot}.
  The dashed line in the left-hand plot traces the relation
  $M \propto \Sigma_t^{3/4}$, which defines a locus of constant
  evaporation time for 
  $\mu_{\rm ev} \propto \Sigma_t^{3/4}$.
\label{fig:sigt}}
\end{figure*}

Figure \ref{fig:sigt}, which should be compared to Figures
\ref{fig:rhoh} and \ref{fig:rhot} above, shows that the average
Galactic GC mass increases systematically with $\Sigma_t$; that the
lower envelope of the $M$--$\Sigma_t$ distribution is described well
by $M\propto \Sigma_t^{3/4}$ (the dashed line in the left-hand panel
of Figure \ref{fig:sigt}), which is a locus of constant lifetime
against evaporation for $\mu_{\rm ev} \propto \Sigma_t^{3/4}$;
and that the scatter in the distribution of cluster $\Sigma_t$ versus
Galactocentric radius (right-hand panel of the figure) is substantial,
as required to account for the almost non-existent correlation between
$M$ and $r_{\rm gc}$.

\begin{figure*}
\centerline{\hfil
    \rotatebox{270}{\includegraphics[height=175mm]{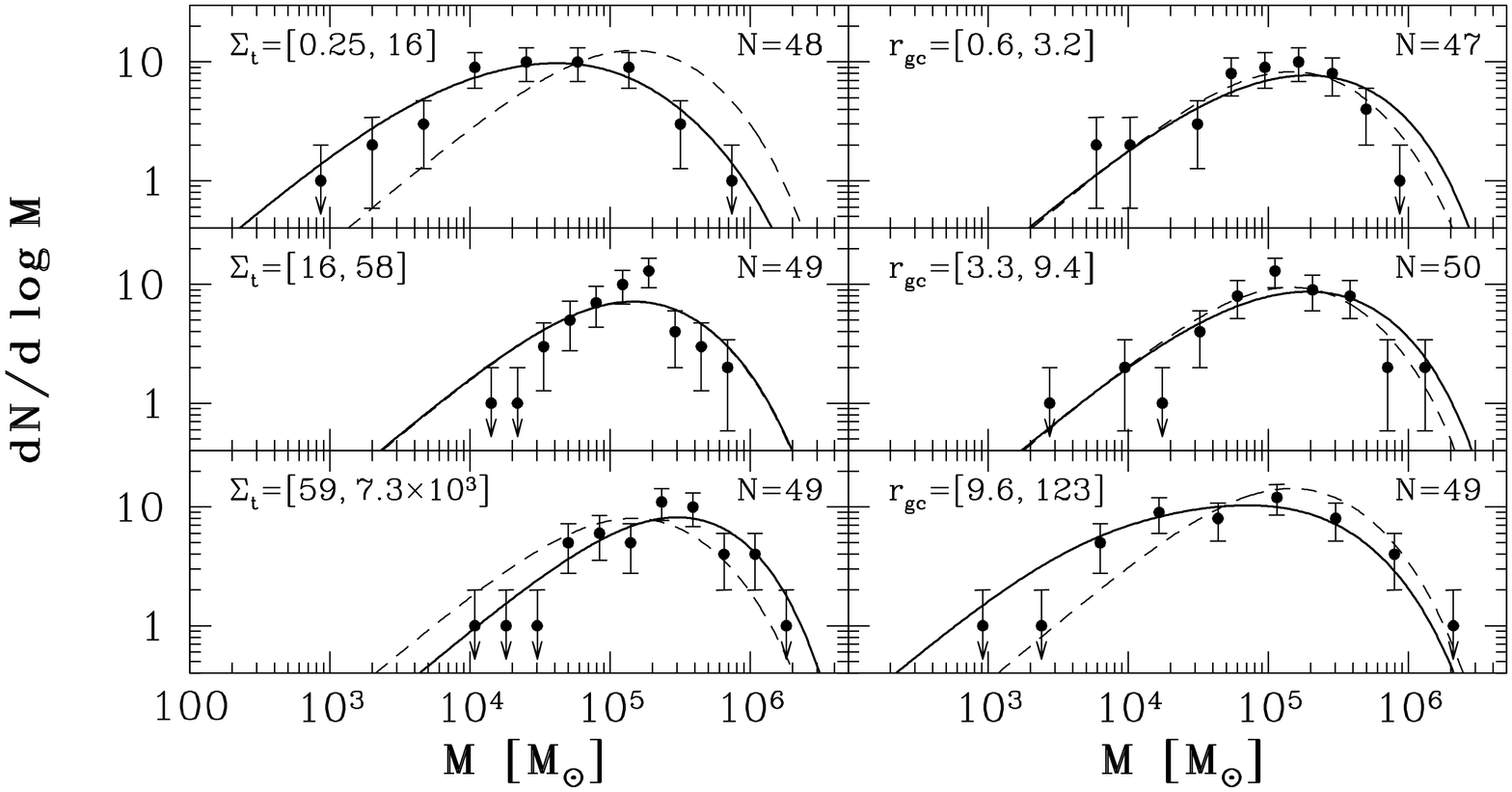}}
\hfil}
\caption{
  Observed GCMF (points, with Poisson errorbars) and models (curves)
  as a function of mean {\it surface} density inside the tidal radius,
  $\Sigma_t \equiv M/\pi r_t^2$ (left-hand panels), and as a
  function of Galactocentric radius, $r_{\rm gc}$ (right-hand
  panels). The dashed curve in every panel is an evolved
  \citeauthor{schechter76} function representing the entire GC system:
  equation (\ref{eq:compschec}) with $\beta=2$, $M_c=10^6\,M_\odot$,
  and a single $\Delta$, common to all clusters, evaluated from
  equation (\ref{eq:deltasig}) using the median $\widehat{\Sigma}_t$
  of all 146 Galactic GCs. Solid curves are subsample-specific models
  using equation (\ref{eq:compschec}) with $\beta=2$ and
  $M_c=10^6\,M_\odot$ but a different $\Delta$ value for every cluster
  (obtained from equation [\ref{eq:deltasig}] using individual
  observational estimates of $\Sigma_t$) in any $\Sigma_t$ or
  $r_{\rm gc}$ bin. 
\label{fig:sigt_gclf}}
\end{figure*}

The left-hand side of Figure \ref{fig:sigt_gclf} shows the mass
functions of globulars in three bins of $\Sigma_t$, as defined 
in each panel. The right-hand side of the
figure shows $dN/d\,\log\,M$ in the same three intervals of
$r_{\rm gc}$ as
in Figures \ref{fig:gclf} and \ref{fig:rhot_gclf} above. As in those
earlier plots, the dashed curve in all panels of Figure
\ref{fig:sigt_gclf} is a model GCMF with the same parameters in every
case, representing the mass function of the entire Galactic GC
system. Once again, compared to the average $M_{\rm TO}$,
the observed turnover mass is significantly lower for clusters in
the lowest $\Sigma_t$ bin and higher for clusters in the highest
$\Sigma_t$ bin, while the width of $dN/d\,\log\,M$ decreases
noticeably as $\Sigma_t$ increases.

The solid curves in Figure \ref{fig:sigt_gclf} are again different in
every panel. They are the sums of evaporation-evolved
\citeauthor{schechter76} functions as in equation (\ref{eq:compschec}),
with the usual $\beta=2$ assumed but with total mass
losses estimated individually for each GC in any $\Sigma_t$ or
$r_{\rm gc}$ bin according to $\Delta \propto \Sigma_t^{3/4}$ rather
than $\Delta \propto \rho_h^{1/2}$ or $\Delta \propto
\rho_t^{1/2}$. However, it turns out not to be necessary to
change the {\it normalization} of $\Delta \propto \rho_h^{1/2}$
in equation (\ref{eq:delta}) to achieve good fits to the
observed GCMF as a function of either $\Sigma_t$ or
$r_{\rm gc}$. Thus, in Figure \ref{fig:sigt_gclf} we have simply used
\begin{equation}
  \Delta =
     1.45\times10^4\ M_\odot\ 
     \left(\Sigma_t/M_\odot\,{\rm pc}^{-2}\right)^{3/4} \ .
\label{eq:deltasig}
\end{equation}

\begin{figure}
\centerline{\hfil
    {\includegraphics[width=85mm]{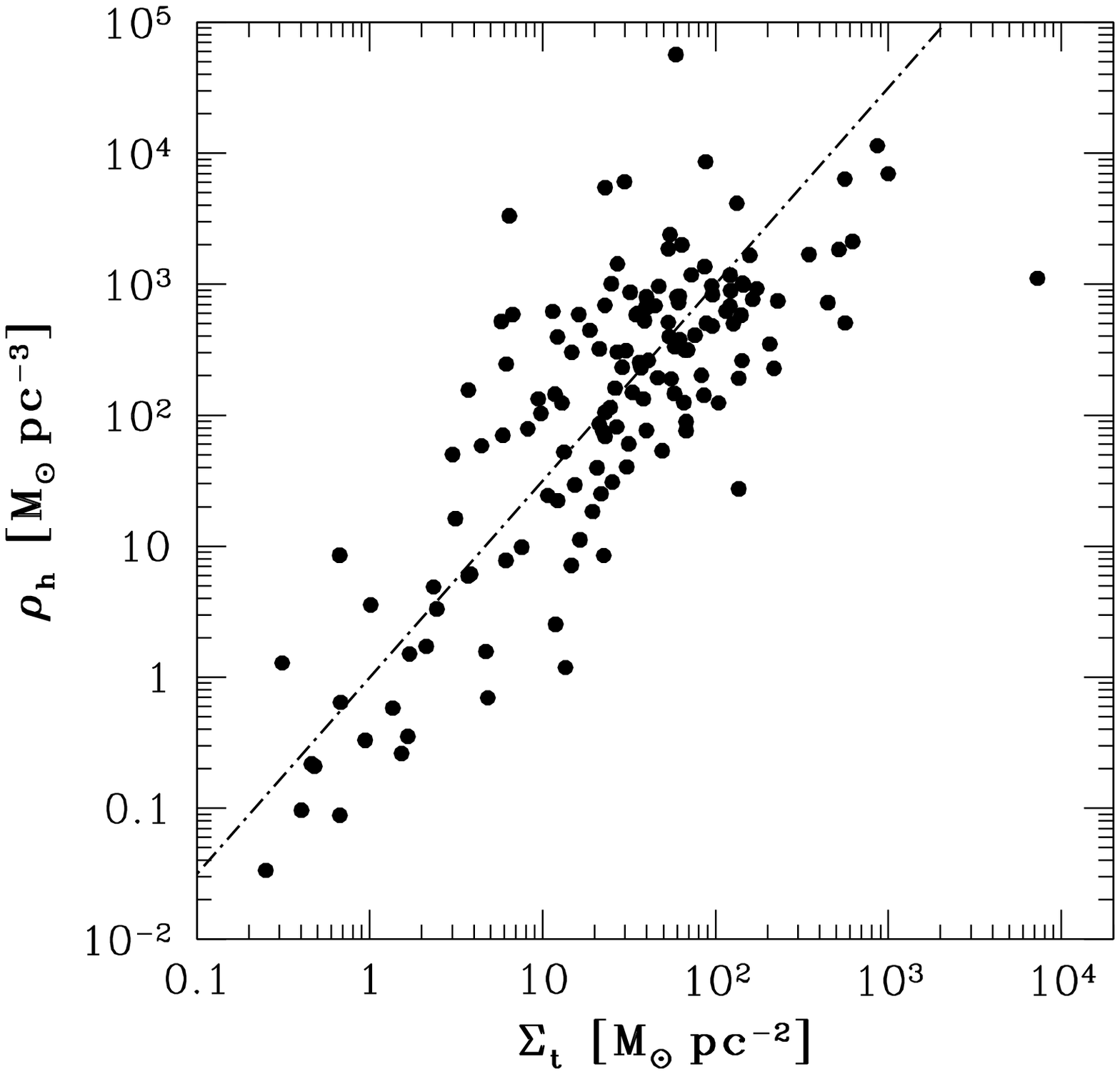}}
\hfil}
\caption{
  Half-mass density, $\rho_h = 3M/8\pi r_h^3$, against mean surface
  density inside the tidal radius, $\Sigma_t = M/\pi r_t^2$, for 146
  clusters with data in \citet{harris96}. The straight line is
  $\rho_h = \Sigma_t^{1.5}$. 
\label{fig:rhoh_sig}}
\end{figure}

The fits of these models, based on $t_{\rm dis} \propto t_{\rm rlx}^x
t_{\rm cr}^{1-x}$ with $x\simeq 0.75$, are indistinguishable from the
fits of our original models based on the standard
$t_{\rm dis} \propto t_{\rm rlx}$, i.e., $x=1$. (We 
have confirmed that adopting individual $\Delta$ given by
equation [\ref{eq:deltasig}] also reproduces the GCMFs of low-and
high-concentration GCs in Figure \ref{fig:conc} as well as before.)
It was somewhat unexpected that equation
(\ref{eq:deltasig}) and equation (\ref{eq:delta}) should have the same
numerical coefficient, but we note that this follows empirically from
the fact that the measured $\rho_h$ and $\Sigma_t$ of Galactic GCs are
consistent with the simple near-equality,
$\rho_h/M_\odot\,{\rm pc}^{-3} \approx
    (\Sigma_t/M_\odot\,{\rm pc}^{-2})^{1.5}$
in the mean. This is illustrated in Figure \ref{fig:rhoh_sig}, which
also shows that there is significant scatter about the
relation.\footnotemark
\footnotetext{Although it may be only a coincidence that the
  constant of proportionality in $\rho_h \propto \Sigma_t^{1.5}$ is so
  near unity, the basic scaling itself holds because combining the
  observed correlation between cluster mass and central concentration
  \citep{djorg94,mcl00} with the intrinsic dependence of
  $r_t/r_h$ on $c$ in \citeauthor{king66} models leads roughly to
  $(r_t/r_h) \propto M^{1/6}$.}
However, this scatter does not correlate with cluster mass or
Galactocentric radius. From a pragmatic point of 
view, therefore, $\rho_h^{1/2}$ and $\Sigma_t^{3/4}$ are near enough
to interchangeable for our purposes, and there is no practical
difference between GCMF models based on one or the other measure of GC
density.

One further check on this is to verify that the mass-loss
rate associated with equation (\ref{eq:deltasig}) is roughly in
keeping with that implied by the $N$-body simulations pointing to
$x=0.75$ in the first place. Thus, we compare the rate
\begin{equation}
\mu_{\rm ev} = \Delta/(13\ {\rm Gyr}) \simeq
        1100\ M_\odot\,{\rm Gyr}^{-1}\ 
        (\Sigma_t/M_\odot\,{\rm pc}^{-2})^{3/4}
\label{eq:muevsig}
\end{equation}
to a formula implicit in
\citetalias{baumgardt03}. Starting with their equation (7) for
the lifetime $t_{\rm dis}$ as a function of initial cluster mass
and perigalactic distance and circular speed in a logarithmic halo
potential; using their $x=0.75$ and their normalization of 
$1.91 \times 10^6$~yr, multiplied as in their equation (9) by
$(1+e)$ to allow for eccentric orbits with apo- and perigalactic
distances related by $e \equiv (r_a-r_p)/(r_a+r_p)$;
inserting their equation (1) for $r_t$; taking the 
mean mass of cluster stars to be $m_*=0.55\, M_\odot$, as
they do; using $\gamma=0.02$ as they do in the Coulomb logarithm,
$\ln(\gamma M_0/m_*)$; and defining
$\Sigma_{t,0} \equiv M_0/\pi r_{t,0}^2$ (the subscript 0 denoting
initial values), we obtain  
\begin{eqnarray}
\mu_{\rm ev}({\rm BM03}) & \simeq & \frac{0.7 M_0}{t_{\rm dis}}
                           \simeq  \frac{560}{1+e}
                                   \ M_\odot\,{\rm Gyr}^{-1}
                                   \nonumber \\
                      ~  &      ~ & \,\, \times \,\,
    \ \left[\frac{\ln \left( 0.036\,M_0/M_\odot \right)}
               {\ln \left( 0.036\times 10^5 \right)}\right]^{3/4}
    \ \left(\frac{\Sigma_{t,0}}{M_\odot\,{\rm pc}^{-2}}\right)^{3/4}
    \ . \nonumber \\
 ~ & ~ & ~
\label{eq:muev_bm03}
\end{eqnarray}
This is appropriate for clusters that just fill their Roche lobes at
perigalacticon, which is where $\Sigma_{t,0}$ is specified.
The factor of 0.7 in the first equality accounts for mass loss
due to stellar evolution in the \citetalias{baumgardt03}
simulations, which, as they discuss, can be treated as having occurred
almost immediately and in full at the beginning of a cluster's life.

Our GCMF-based $\mu_{\rm ev}$ is a factor of $\approx\! 2$ faster than
the $N$-body value for clusters on circular orbits (with $e=0$ and in
steady tidal fields) in the simulations; and our $\mu_{\rm ev}$ is
still within a factor of about three of the $N$-body rate for clusters
on eccentric orbits with $e = 0.5$ in \citetalias{baumgardt03}
($e\simeq 0.5$--0.6 is typical for tracers with an isotropic velocity
distribution in a logarithmic potential; \citealt{vandenbosch99}).
This is very similar to the comparison of lifetimes in
\S\ref{subsec:lifetime} for our original models based on
$\mu_{\rm ev} \propto \rho_h^{1/2}$. Moreover, our new estimate of
$\mu_{\rm ev}$ and that in \citetalias{baumgardt03} are still subject to
their own, separate uncertainties and reflect different idealizations
and assumptions. For example, our rate still depends on the exact
power-law exponent $\beta$ at low masses in the initial GCMF, as discussed
after equation (\ref{eq:tdis_scale}); while the rate from
\citetalias{baumgardt03} still neglects gravitational shocks from disk
crossings and passages by a discrete galactic bulge, and may
additionally be biased low for $M_0 > 10^5\,M_\odot$ if $x>0.75$ at
such masses. All of this---not to mention again the large
uncertainties and possible systematics in the estimates of tidal radii
needed to calculate $\Sigma_t$---makes the near agreement between
equations (\ref{eq:muevsig}) and (\ref{eq:muev_bm03}) more striking
than any apparent discrepancy.

In summary, although the relation
$\mu_{\rm ev} \propto \rho_h^{1/2} \simeq {\rm constant}$ in time
is rigorously correct only in rather specific circumstances, our
GCMF models based on it in \S\ref{sec:results} are good proxies, in
all respects, for models based on other plausible characterizations of
relaxation-driven cluster mass loss. This result will likely be
important for future studies of the mass functions of extragalactic
cluster systems, where it may well be necessary to adopt procedures
based on $\rho_h$ rather than $\rho_t$ or $\Sigma_t$ because of
the difficulty or impossibility of estimating tidal radii.

\subsection{$M_{\rm TO}$ versus $r_{\rm gc}$, and
            Velocity Anisotropy in GC Systems}
\label{subsec:anisotropy}

In this paper we have directly modeled $dN/d\,\log\,M$ as a function
only of GC density and age, and used the observed $\rho_h$ (or
$\rho_t$, or $\Sigma_t$) of clusters
in relatively narrow ranges of Galactocentric position to show that
such models are consistent with the current near-constancy of
the GCMF as a function of $r_{\rm gc}$. Most other
models in the literature for evaporation-dominated GCMF evolution,
in either the Milky Way or other galaxies, instead predict the
distribution explicitly as a function of $r_{\rm gc}$
at any time. They therefore need, in effect, to derive 
theoretical density--position relations for clusters in galaxies
alongside their main GCMF calculations. This usually begins with the
adoption of analytical potentials to describe the parent 
galaxies of GCs. Taking these to be spherical and static
for a Hubble time allows the use of standard tidal-limitation
formulae to write GC densities ab initio in terms of the
(fixed) pericenters $r_p$ of unique orbits in the adopted potentials.
Cluster relaxation times and mass-loss rates $\mu_{\rm ev}$ then
follow as functions of $r_p$ as well. Finally, specific initial mass,
space, and velocity (or orbital eccentricity) distributions are chosen
for entire GC systems, so that at all later times it is known what the
dynamically evolved $dN/d\,\log\,M$ is for globulars with any single
$r_p$; how many clusters with a given $r_p$ survive; and what the
distributions of $r_p$ and all dependent cluster properties are at any
instantaneous position $r_{\rm gc}$.

In this approach, if the GCMF began with a power-law rise towards low
masses and its current peak is due entirely to cluster disruption,
then a dependence of $M_{\rm TO}$ on $r_p$ is expected in general,
because the densities of tidally limited GCs decrease with increasing
$r_p$. Thus, models along these lines that assume the orbit
distribution of a GC system to be the same at all radii in a galaxy 
(i.e., that the time average of the ratio $r_{\rm gc}/r_p$
is independent of position) have typically had difficulty in
accounting for the observed weak or non-correlation between
$M_{\rm TO}$ and present $r_{\rm gc}$ in large galaxies. This is
particularly a problem if it is assumed that the initial GCMF was a
pure power law, with the same index at arbitrarily high masses as low
(e.g., \citealt{baumgardt98}; \citealt{vesperini01}). It is
potentially less of a concern if $dN/d\,\log\,M$ started as a
\citeauthor{schechter76} function with an exponential cut-off at
masses $M>M_c$, as we have assumed, since then the existence of a
strict upper bound $M_{\rm TO} \le M_c$ (\S\ref{subsec:models}) means
that the dependence of an evaporation-evolved $M_{\rm TO}$ on $r_p$
and $r_{\rm gc}$ must saturate for small enough galactocentric radii
(high enough GC densities). Even so, the ``scale-free'' models of
\citetalias{fall01}, in which $M_c \simeq 10^6\,M_\odot$ and all GCs
in a Milky Way-like galaxy potential have the same time-averaged
$r_{\rm gc}/r_p$, still predict a gradient in $M_{\rm TO}$ versus
$r_{\rm gc}$ that is stronger than observed.

\citetalias{fall01} showed that, if they left all of their other
assumptions unchanged, then a dependence of GCMF peak mass on
$r_{\rm gc}$ could be effectively erased by an appropriately varying
radial velocity anisotropy in the initial GC system. Thus, in their 
``Eddington'' models the eccentricity of a typical cluster orbit
increases with galactocentric distance (the time average of
$r_{\rm gc}/r_p$ increases with radius), such that globulars spread
over a larger range of current $r_{\rm gc}$ can have more similar
$r_p$ and associated $M_{\rm TO}$. However, the initial
velocity-anisotropy gradient required to fit the Milky Way GCMF data
specifically is only marginally consistent with the observed
kinematics of the GC system \citep[e.g.,][]{dinescu99}.\footnotemark
\footnotetext{The fact that clusters on radial orbits are
  preferentially disrupted lessens any inconsistency between the
  radial anisotropy required in the initial velocity distribution and
  observational constraints on the present velocity distribution.}
Subsequently,
\citet{vesperini03b} constructed broadly similar models for the
GCMF of the Virgo elliptical M87 and concluded that there, too, a
variable radial velocity anisotropy is required to match the observed
$M_{\rm TO}$ versus $r_{\rm gc}$; but the model anisotropy profile in
this case is clearly inconsistent with the true velocity
distribution of the GC system, which is observed to be isotropic out
to large $r_{\rm gc}$ \citep{romanowsky01,cote01}.

These results certainly suggest that some element is lacking in
$r_{\rm gc}$-oriented GCMF models developed as outlined
above. But they do not mean that the fault lies with the main
hypothesis, that the difference between the mass functions of young
clusters and old GCs is due to the effects of slow,
relaxation-driven disruption in the latter case. Any conclusions about
velocity anisotropy depend on the totality of steps taken
to connect the densities and positions of clusters; and it is possible
that reasonable changes to one or more of these ancillary assumptions
could make the models compatible with the observed kinematics of GCs
in both the Milky Way and M87, without abandoning a basic physical
picture of evaporation-dominated GCMF evolution that is otherwise
quite successful. 

One issue is that previous models have always specified evaporation
rates a priori as functions of cluster density (or orbital
pericenter), usually normalizing $\mu_{\rm ev}$ so that
$t_{\rm dis}/t_{\rm rh} \simeq 20$--40 as in standard treatments of
two-body relaxation. However, following our
discussion in \S\ref{subsec:lifetime} and \S\ref{subsec:approx}, it
would seem worthwhile to investigate these models with $\mu_{\rm ev}$
increased at fixed $\rho_h$ or $r_p$ to allow
$t_{\rm dis}/t_{\rm rh} \approx 10$ (if $\beta\simeq 2$ for the
low-mass power-law part of the initial GCMF).

\citetalias{fall01} and \citet{vesperini03b} both consider
velocity distributions parametrized by a galactocentric anisotropy
radius, $R_A$, inside of which a cluster system is essentially
isotropic and beyond which it is increasingly dominated by radial
orbits. In these terms, the difficulty with the published models is
that, to reproduce the observed insensitivity of $M_{\rm TO}$ to
$r_{\rm gc}$ given standard normalizations of $\mu_{\rm ev}$, they
require values of $R_A$ that are smaller than allowed by
observations (especially for M87). Increasing $R_A$ to more realistic
values while keeping 
the normalization of $\mu_{\rm ev}$ fixed leads to a stronger gradient
in $M_{\rm TO}$: the orbits of GCs at small $r_{\rm gc} \la R_A$
remain closely isotropic and the typical $r_p$ and $M_{\rm TO}$ are
essentially unchanged, while at large galactocentric distances the
cluster orbits are on average less radial than before,
with larger $r_p$, lower densities, and lower evolved $M_{\rm TO}$ for
a given $r_{\rm gc}$. This effect is illustrated, for example, in
Figure 9 of \citetalias{fall01}. However, it can be compensated at
least in part by increasing $\mu_{\rm ev}$ by a common factor for
all GCs, with the new, larger $R_A$ fixed, {\it if the initial mass
function is assumed to have been a \citeauthor{schechter76} function
rather than a pure power law extending to arbitrarily high masses}. A
faster evaporation rate will then lead to a (roughly) proportionate
increase in the evolved GCMF peak mass for GCs with relatively low
densities, i.e., those at large $r_{\rm gc}$ and $r_p$; but the
increase in $M_{\rm TO}$ will be {\it smaller}, and eventually even
negligible, for higher-density clusters at progressively smaller
$r_{\rm gc}$---again because $M_{\rm TO}$ grows less than linearly
with $\mu_{\rm ev} \propto \rho_h^{1/2}$ when there is an upper limit
$M_{\rm TO} < M_c$ due to an exponential cut-off in the initial
$dN/d\,\log\,M_0$. Thus, the  
qualitative effect of increasing the normalization of $\mu_{\rm ev}$
in models with radially varying GC velocity anisotropy is to weaken
the amount of radial-orbit bias required to fit an observed
$M_{\rm TO}$ versus $r_{\rm gc}$.

Another point, emphasized by \citetalias{fall01}, has to do with the
standard starting assumption that GCs orbit in galaxies that are
perfectly static and spherical. In reality, galaxies grow
hierarchically. In this case, even if the values of
$\mu_{\rm ev}$ are not changed, much of the burden for the weakening or
erasing of any initial gradients in $M_{\rm TO}$ versus $r_{\rm gc}$
may be transferred from velocity anisotropy to the time-dependent
evolution of the galaxies themselves. Violent relaxation, major
mergers, and smaller accretion events all work to move clusters between
different parts of galaxies and between different progenitors,
scrambling and combining any number of
pericenter--density--$M_{\rm TO}$ relations.
Any position dependences in the GC $\rho_h$ distribution and in
$M_{\rm TO}$ itself for the final galaxy are therefore bound to be
weaker, more scattered, and more difficult to relate accurately to a
cluster velocity distribution than in the case of a monolithic,
non-evolving potential. Allowing for a non-spherical galaxy potential
would have qualitatively the same effect, because in this case every
cluster explores a range of pericenters and different maximum tidal
fields on each of its orbits.

In this situation, it may be important to ask how evaporation
rates can still be approximately constant in time---so
that cluster masses still decrease approximately linearly with $t$ as
our models assume---if the tidal field around any given GC
changes significantly over time. Thus,
consider first a system of GCs in a single, static galaxy
potential. The mass-evolution curve for each cluster is approximately
a straight line,
$M(t) \simeq M_0 - \mu_{\rm ev} t$, with $\mu_{\rm ev}$ depending on 
some measure of internal density, which may be $\rho_h^{1/2}$,  
$\rho_t^{1/2}$, or $\Sigma_t^{3/4}$. The average mass-evolution curve 
for the entire system of clusters is also approximately linear,
$\langle M(t) \rangle \simeq \langle M_0 \rangle -
                             \langle \mu_{\rm ev} \rangle t$.
If now a merger or other event rearranges the clusters in the galaxy,
then after the event the mass-loss rates of some clusters will be
higher than before and the rates of other clusters will be lower than
before. However, if the mean density of the galaxy
as a whole is roughly the same after the event as before, then so
too will be the average of the GC densities, because of tidal
limitation. The average
$\langle \mu_{\rm ev} \rangle \propto \langle \rho_h^{1/2} \rangle$
(say) will differ even less between the pre- and post-merger
systems. Thus, although using instantaneous densities to estimate the
past $\mu_{\rm ev}$ of individual clusters may err on the high side for
some clusters and on the low side for others, these errors will
average away to a small or even zero net bias. The approximation
$\mu_{\rm ev}\simeq {\rm constant}$~in time in our GCMF models
will then still be valid in the mean, and the average
$\langle M(t) \rangle$ dependence of sufficiently large numbers of
clusters will remain roughly linear. 

This type of scenario might be expected to pertain at least to galaxies
that evolve on the fundamental plane, since this entails a connection
between the total (baryonic plus dark) masses and circular speeds of
galaxies, of the form $M_{\rm gal} \propto V_c^3$ or
$M_{\rm gal}\propto V_c^4$. By the virial theorem, the average
densities scale as  $\rho_{\rm gal} \propto V_c^6/M^2$, and thus
$\rho_{\rm gal} \propto M_{\rm gal}^0$ or
$\rho_{\rm gal} \propto M_{\rm gal}^{-1/2}$.
Insofar as
$\langle \rho_h \rangle \propto \langle \rho_t \rangle
         \propto \rho_{\rm gal}$
for the GCs, the system-wide average
$\langle \mu_{\rm ev} \rangle \propto \langle \rho_h^{1/2} \rangle$
should therefore not change drastically even after a major merger
between two fundamental-plane galaxies; at
most, the ratio of final to initial $\langle \mu_{\rm ev} \rangle$
will be roughly of order the $-1/4$ power of the ratio of final to
initial $M_{\rm gal}$. Note that this line of reasoning is closely
related to that applied by \citetalias{fall01} to explain the small
observed galaxy-to-galaxy differences in the average turnover
masses of entire GC systems (although non-zero differences do exist,
and can be accomodated in these sorts of arguments; see
\citealt{jordan06,jordan07}).

A full exploration of questions such as these, about the wide range of
ingredients in current GC-plus-galaxy models, will most likely require
large $N$-body simulations set in a realistic, cold dark matter
cosmology. Until these can be carried out, it is our view that the
kinematics of globular cluster systems cannot be used as decisive side
constraints on theories for the GCMF.

\section{Conclusions}
\label{sec:conclusions}

We have shown that the mass function $dN/d\,\log\,M$ of globular
clusters in the Milky Way depends significantly on cluster half-mass
density, $\rho_h$, with the peak or turnover mass $M_{\rm TO}$
increasing and the width of the distribution decreasing as $\rho_h$
increases. This behavior is expected if the GCMF initially
rose towards masses below the present turnover scale---as the mass
functions of young cluster systems like that in the Antennae galaxies
do---and has evolved to its current shape via the slow depletion of
low-mass clusters over Gyr timescales, primarily through
relaxation-driven evaporation. The fact that $M_{\rm TO}$
{\it increases} with cluster density favors evaporation over external
gravitational shocks as the primary mechanism of low-mass cluster
disruption, since the mass-loss rates associated with shocks depend
inversely on cluster density and directly on cluster mass. Our results
therefore add to previous arguments supporting an interpretation of
the GCMF in terms of evaporation-dominated evolution, based on the
fact that $dN/d\,\log\,M$ scales as $M^{1-\beta}$ with
$\beta \simeq 0$ in the low-mass limit \citep{fall01}.

The observed GCMF as a function of $\rho_h$ is fitted well by simple
models in which the initial distribution was a
\citeauthor{schechter76} function,
$dN/d\,\log\,M_0 \propto M_0^{1-\beta}\,\exp\,\left(-M_0/M_c\right)$
with $\beta=2$ and $M_c \simeq 10^6\,M_\odot$ assumed, and in which
clusters have been losing mass for a Hubble time at roughly steady
rates that can be estimated from their current half-mass densities as
$\mu_{\rm ev} \propto \rho_h^{1/2}$. We have shown that, although
this prescription is approximate, it captures
the main physical dependence of relaxation-driven evaporation.
In particular, it leads to model GCMFs that are entirely consistent
with those resulting from alternative characterizations of 
evaporation rates in terms of cluster tidal densities $\rho_t$ or mean
surface densities $\Sigma_t$ (\S\ref{subsec:approx}).
The normalization of $\mu_{\rm ev}$ at a given $\rho_h$ (or $\rho_t$,
or $\Sigma_t$) required to fit the GCMF implies total cluster
lifetimes that are within range of the lifetimes typically obtained in
theoretical studies of two-body relaxation, although our values may be
slightly shorter than the theoretical ones if the low-mass, power-law
part of the initial cluster mass function was as steep as we have
assumed.

Taking clusters in various bins of central concentration $c$ and
Galactocentric radius $r_{\rm gc}$ and using their (individual)
observed densities as direct input to our models yields dynamically
evolved GCMFs as functions of $c$ and $r_{\rm gc}$ that agree well
with all data. This again indicates that the most fundamental physical
dependence in the GCMF is that on cluster density. Moreover,
our models for $dN/d\,\log\,M$ versus $r_{\rm gc}$ obtained in this
way are consistent in particular with the well-known insensitivity of
the GCMF peak mass to Galactocentric position. This is seen to
follow from a significant variation of $M_{\rm TO}$ with $\rho_h$ (or
$\rho_t$, or $\Sigma_t$)---due in our analysis to
evaporation-dominated cluster disruption---combined with substantial
scatter in the GC densities at any Galactocentric position.

We have not invoked an anisotropic GC velocity distribution to explain
the observed weak variation of $M_{\rm TO}$ with $r_{\rm gc}$; indeed,
we have made no predictions or assumptions whatsoever about velocity
anisotropy. We have emphasized that, when velocity anisotropy enters
other long-term dynamical-evolution models for the GCMF, it is only in
conjunction with several additional, interrelated assumptions made as
part of larger efforts to derive theoretical density--$r_{\rm gc}$
relations for GCs---which we have not attempted to do here. The
apparent need in some current models for a strong bias towards
high-eccentricity cluster orbits to explain the near-constancy of
$M_{\rm TO}$ versus $r_{\rm gc}$ might well be avoided by changing one
or more ancillary assumptions in the models, without having to discard
the underlying idea that the peak and low-mass shape of the GCMF are
the result of relaxation-driven cluster disruption.

It clearly will be of interest to test and refine the main ideas in
this paper through modeling of the GCMFs in other galaxies.
For the time being at least, doing so will require the estimation of
approximate mass-loss rates using cluster half-mass densities
rather than tidal quantities, simply because GC half-light radii can
be measured accurately in many systems beyond the Local
Group, whereas tidal radii are much more model-dependent and difficult
to observe. \citet{chandar07} have recently shown that
the peak mass of the GCMF in the Sombrero galaxy (M104) increases  
with $\rho_h$ in a way that is reasonably well described by sums of
evolved \citet{schechter76} functions as in the models presented in
this paper. It should be relatively straightforward to pursue similar
studies in other nearby galaxies.

\acknowledgements

We thank Michele Trenti, Douglas Heggie, Bill Harris, Rupali
Chandar, and Bruce Elmegreen for helpful discussions and comments.   
SMF acknowledges support from the Ambrose Monell Foundation and from
NASA grant AR-09539.1-A, awarded by the Space Telescope Science
Institute, which is operated by AURA, Inc., under NASA contract
NAS5-26555.


\begin{thebibliography}{}

\bibitem[Aguilar, Hut, \& Ostriker(1988)]{aguilar88} Aguilar, L., Hut,
  P., \& Ostriker, J. P. 1988, \apj, 335, 720

\bibitem[Barmby, Huchra, \& Brodie(2001)]{barmby01} Barmby, P., Huchra, J. P.,
  \& Brodie, J. P. 2001, \aj, 121, 1482

\bibitem[Barmby et al.(2007)]{barmby07} Barmby, P., McLaughlin, D. E., Harris,
  W. E., Harris, G. L. H., \& Forbes, D. A. 2007, \aj, 133, 2764

\bibitem[Baumgardt(1998)]{baumgardt98} Baumgardt, H. 1998, \aap, 330, 480

\bibitem[Baumgardt(2001)]{baumgardt01} Baumgardt, H. 2001, \mnras,
  325, 1323

\bibitem[Baumgardt \& Makino(2003)]{baumgardt03} Baumgardt, H., \& Makino,
  J. 2003, \mnras, 340, 227 (\citetalias{baumgardt03})

\bibitem[Binney \& Tremaine(1987)]{bt87} Binney, J., \& Tremaine, S. 1987,
  Galactic Dynamics (Princeton: Princeton University Press)

\bibitem[Burkert \& Smith(2000)]{burkert00} Burkert, A., \& Smith, G. H. 2000,
  \apj, 542, L95

\bibitem[Caputo \& Castellani(1984)]{caputo84} Caputo, F., \&
  Castellani, V. 1984, \mnras, 207, 185

\bibitem[Chandar, Fall, \& McLaughlin(2007)]{chandar07} Chandar, R.,
  Fall, S. M., \& McLaughlin, D. E. 2007, \apj, 668, L119

\bibitem[Chandrasekhar(1942)]{chandra42} Chandrasekhar, S. 1942,
  Principles of Stellar Dynamics (Chicago: University of Chicago
  Press)

\bibitem[Chernoff \& Weinberg(1990)]{chernoff90} Chernoff, D. F., \&
  Weinberg, M. D. 1990, \apj, 351, 121

\bibitem[C\^ot\'e et al.(2001)]{cote01} C\^ot\'e, P., et~al. 2001,
  \apj, 559, 828 

\bibitem[Dinescu, Girard, \& van Altena(1999)]{dinescu99} Dinescu, D. I.,
  Girard, T. M., \& van Altena, W. F. 1999, \aj, 117, 1792

\bibitem[Djorgovski \& Meylan(1994)]{djorg94} Djorgovski, S., \&
  Meylan, G. 1994, \aj, 108, 1292

\bibitem[Elmegreen \& Efremov(1997)]{elmegreen97} Elmegreen, B. G., \&
  Efremov, Y. N. 1997, \apj, 480, 235

\bibitem[Fall \& Rees(1977)]{fall77} Fall, S. M., \& Rees, M. J. 1977, \mnras,
  181, 37P

\bibitem[Fall \& Zhang(2001)]{fall01} Fall, S. M., \& Zhang, Q. 2001, \apj,
  561, 751 \citepalias{fall01}

\bibitem[Fukushige \& Heggie(2000)]{fukushige00} Fukushige, T., \&
  Heggie, D. C. 2000, \mnras, 318, 753

\bibitem[Giersz(2001)]{giersz01} Giersz, M. 2001, \mnras, 324, 218

\bibitem[Giersz \& Heggie(1996)]{giersz96} Giersz, M., \& Heggie,
   D. C. 1996, \mnras, 279, 1037

\bibitem[Gnedin(1997)]{gnedin97} Gnedin, O. Y. 1997, \apj, 487, 663

\bibitem[Gnedin \& Ostriker(1997)]{gnedost97} Gnedin, O. Y., \& Ostriker,
  J. P. 1997, \apj, 474, 223

\bibitem[Gnedin, Lee, \& Ostriker(1999)]{gnedin99} Gnedin, O. Y., Lee, H. M.,
  \& Ostriker, J. P. 1999, \apj, 522, 935

\bibitem[Harris(1996)]{harris96} Harris, W. E. 1996, \aj, 112, 1487

\bibitem[Harris(2001)]{harris01} Harris, W.E. 2001, in Star Clusters (28th
  Saas-Fee Advanced Course) ed. L. Labhardt \& B. Binggeli (Berlin: Springer),
  223

\bibitem[Harris \& Pudritz(1994)]{harris94} Harris, W. E., \& Pudritz,
  R. E. 1994, \apj, 429, 177

\bibitem[Harris, Harris, \& McLaughlin(1998)]{harris98} Harris, W. E., Harris,
  G. L. H., \& McLaughlin, D. E. 1998, \aj, 115, 1801

\bibitem[H\'enon(1961)]{henon61} H\'enon, M. 1961, Ann. d'Astrophys.,
  24, 369

\bibitem[Innanen, Harris, \& Webbink(1983)]{innanen83} Innanen, K. A.,
  Harris, W. E., \& Webbink, R. F. 1983, \aj, 88, 338

\bibitem[Johnstone(1993)]{johnstone93} Johnstone, D. 1993, \aj, 105, 155

\bibitem[Jord\'an et al.(2005)]{jordan05} Jord\'an, A., et al. 2005, \apj,
  634, 1002

\bibitem[Jord\'an et al.(2006)]{jordan06} Jord\'an, A., et al. 2006, \apj,
  651, L25 

\bibitem[Jord\'an et al.(2007)]{jordan07} Jord\'an, A., et al. 2007, \apjs, 
  171, 101

\bibitem[Joshi, Nave, \& Rasio(2001)]{joshi01} Joshi, K. J., Nave,
  C. P., \& Rasio, F. A. 2001, \apj, 550, 691

\bibitem[Kavelaars \& Hanes(1997)]{kavelaars97} Kavelaars, J. J., \& Hanes,
  D. A. 1997, \mnras, 285, L31

\bibitem[Lee \& Goodman(1995)]{lee95} Lee, H. M., \& Goodman, J. 1995, \apj,
  443, 109

\bibitem[King(1958)]{king58} King, I.  1958, \aj, 63, 109

\bibitem[King(1959)]{king59} King, I.  1959, \aj, 64, 351

\bibitem[King(1966)]{king66} King, I. R.  1966, \aj, 71, 64

\bibitem[Lee \& Ostriker(1987)]{lee87} Lee, H. M., \& Ostriker,
  J. P. 1987, \apj, 322, 123

\bibitem[McLaughlin(2000)]{mcl00} McLaughlin, D. E. 2000, \apj, 539, 618

\bibitem[McLaughlin \& van der Marel(2005)]{mcl05} McLaughlin, D. E., \& van
  der Marel, R. P. 2005, \apjs, 161, 304

\bibitem[Murali \& Weinberg(1997)]{murali97} Murali, C., \& Weinberg,
  M. D. 1997, \mnras, 291, 717

\bibitem[Okazaki \& Tosa(1995)]{okazaki95} Okazaki, T., \& Tosa,
  M. 1995, \mnras, 274, 48

\bibitem[Ostriker \& Gnedin(1997)]{ostriker97} Ostriker, J. P., \&
  Gnedin, O. Y. 1997, \apj, 487, 667

\bibitem[Parmentier \& Gilmore(2007)]{parmentier07} Parmentier, G., \&
  Gilmore, G. 2007, \mnras, 377, 352

\bibitem[Prieto \& Gnedin(2006)]{prieto06} Prieto, J. L., \& Gnedin,
  O. Y. 2006, preprint ({\tt astro-ph/0608069})

\bibitem[Romanowsky \& Kochanek(2001)]{romanowsky01} Romanowsky,
  A. J., \& Kochanek, C. S. 2001, \apj, 553, 722

\bibitem[Schechter(1976)]{schechter76} Schechter, P. 1976, \apj, 203, 297

\bibitem[Smith \& Burkert(2002)]{smith02} Smith, G. H., \& Burkert, A. 2002,
  \apj, 578, L51

\bibitem[Spitler et al.(2006)]{spitler06} Spitler, L. R., Larsen,
  S. S., Strader, J., Brodie, J. P., Forbes, D. A., \& Beasley,
  M. A. 2006, \aj, 132, 1593

\bibitem[Spitzer(1987)]{spitzer87} Spitzer, L. 1987, Dynamical Evolution of
Globular Clusters (Princeton: Princeton Univ. Press)

\bibitem[Takahashi \& Portegies Zwart(1998)]{takahashi98} Takahashi,
  K., \& Portegies Zwart, S. F. 1998, \apj, 503, L49

\bibitem[Takahashi \& Portegies Zwart(2000)]{takahashi00} Takahashi,
  K., \& Portegies Zwart, S. F. 2000, \apj, 535, 759

\bibitem[Trenti, Heggie, \& Hut(2007)]{trenti07} Trenti, M., Heggie,
  D. C., \& Hut, P. 2007, \mnras, 374, 344

\bibitem[van den Bosch et al.(1999)]{vandenbosch99} van den Bosch,
  F. C., Lewis, G. F., Lake, G., \& Stadel, J. 1999, \apj, 515, 50

\bibitem[Vesperini(1997)]{vesperini97a} Vesperini, E. 1997, \mnras,
  287, 915

\bibitem[Vesperini(1998)]{vesperini98} Vesperini, E. 1998, \mnras, 299, 1019

\bibitem[Vesperini(2000)]{vesperini00} Vesperini, E. 2000, \mnras, 318, 841

\bibitem[Vesperini(2001)]{vesperini01} Vesperini, E. 2001, \mnras, 322, 247

\bibitem[Vesperini \& Heggie(1997)]{vesperini97b} Vesperini, E., \& Heggie,
  D. C. 1997, \mnras, 289, 898

\bibitem[Vesperini \& Zepf(2003)]{vesperini03a} Vesperini, E., \& Zepf,
  S. E. 2003, \apj, 587, L97

\bibitem[Vesperini et al.(2003)]{vesperini03b} Vesperini, E., Zepf, S. E.,
  Kundu, A., \& Ashman, K. M. 2003, \apj, 593, 760

\bibitem[Waters et al.(2006)]{waters06} Waters, C. Z., Zepf, S. E.,
  Lauer, T. R., Baltz, E. A., \& Silk, J. 2006, \apj, 650, 885

\bibitem[Zhang \& Fall(1999)]{zhang99} Zhang, Q., \& Fall, S. M. 1999, \apj,
  527, L81

\end{thebibliography}
\end{document}